%% file: main.tex
\documentclass{article}

\usepackage[preprint]{neurips_2026}

\usepackage[utf8]{inputenc}
\usepackage[T1]{fontenc}
\usepackage{hyperref}
\usepackage{url}
\usepackage{booktabs}
\usepackage{amsfonts}
\usepackage{amsmath}
\usepackage{nicefrac}
\usepackage{microtype}
\usepackage{graphicx}
\usepackage{xcolor}
\usepackage{algorithm}
\usepackage{algorithmic}
\usepackage{multirow}
\usepackage{subcaption}
\usepackage{tikz}
\usepackage{pgfplots}
\usepackage{pifont}
\usepackage{enumitem}
\pgfplotsset{compat=1.18}
\usetikzlibrary{patterns, arrows.meta, positioning, calc, decorations.pathreplacing}

\title{PEEK: Predictive Queue-Informed KV Cache Management for LLM Serving\thanks{Code: \url{https://github.com/xiexbing/peek}}}

\author{
  Bing Xie\thanks{Corresponding author.} \and Zhipeng Wang \and Masahiro Tanaka \and Zheng Zhen \\
}

\begin{document}

\maketitle

\input{content/abstract}

\input{content/introduction}

\input{content/background}

\input{content/method}

\input{content/experimental_setup}

\input{content/results}

\input{content/related_work}

\input{content/conclusion}

\setlength{\bibsep}{2pt plus 1pt minus 1pt}
\bibliographystyle{plainnat}
\bibliography{references}

\appendix
\renewcommand{\thesection}{A.\arabic{section}}

\input{content/appendix}

\end{document}

%% file: content/abstract.tex
\begin{abstract}
We present PEEK, a lightweight scheduling and eviction framework for both online (streaming) and offline (batch) LLM serving; this paper focuses on the online regime. PEEK maintains an incremental \emph{radix tree over the pending queue}, exposing prefix-sharing clusters no existing engine surfaces. A low-overhead \emph{dual-walk} matches the tree against the engine's prefix cache to yield longest-prefix-match for every waiting request; PEEK then admits cluster pioneers first so siblings inherit the freshly cached prefix, a co-designed eviction hook protects blocks ancestral to queued demand, and a \emph{multi-lane} stride scheduler bounds starvation. On SGLang and vLLM across five workloads up to 4$\times$H100 (DP=2 over TP=2), PEEK delivers up to $\mathbf{3.0\times/2.6\times}$ cache hit, $\mathbf{7.9\times/7.1\times}$ TTFT, $\mathbf{6.7\times/5.5\times}$ E2E, and $\mathbf{3.6\times/4.5\times}$ throughput gains over each engine's strongest stock baseline (SGLang/vLLM), while matching baselines within noise on workloads with no exploitable prefix structure. Wins hold as KV-cache pressure and inference parallelism scale.
\end{abstract}

%% file: content/introduction.tex
\section{Introduction}

Modern LLM serving engines optimize KV cache reuse: SGLang's RadixAttention~\citep{zheng2024sglang} provides longest-prefix matching over cached blocks; vLLM's PagedAttention~\citep{kwon2023vllm} enables block-level sharing via OS-style paging. Yet both stop at the cache---neither reasons about which \emph{waiting} requests share prefixes with each other, leaving the predictive signal in the request queue unexploited (Table~\ref{tab:policy-comparison}). SGLang's prefix-aware policies (\textbf{LPM}, \textbf{DFS\_WEIGHT}) remain bound to current cache state and degrade once the cache is cold; vLLM defaults to FCFS, which under memory pressure devolves into \emph{cache thrashing} as interleaved prefix groups evict one another (\S\ref{sec:thrashing}). In both engines, scheduling and eviction are decoupled, precluding joint optimization---inefficiencies magnified in long-context workloads such as RAG and agentic pipelines.

We present \textbf{PEEK}, a lightweight scheduling and eviction framework for both online (streaming) and offline (batch) LLM serving; \emph{this paper focuses on online}. Our motivation is a simple observation: \textbf{the waiting queue is itself a structured workload whose prefix-sharing relationships can directly drive KV-cache reuse}. PEEK retains LPM's prefix-match objective but feeds it richer signals at lower overhead from an incremental radix tree via three mechanisms (Appendix Table~\ref{tab:peek-vs-lpm}):

\begin{enumerate}[leftmargin=*,itemsep=1pt,topsep=2pt,parsep=0pt]
    \item \textbf{See the queue's structure for cheap (\S\ref{sec:dual_walk}).} PEEK maintains an incremental radix tree over the pending queue---no per-cycle rebuild, $O(D)$ per insert/remove, $O(1)$ per query on scheduling/eviction signals. To compute LPM hits against the engine cache, PEEK's \emph{dual-walk} co-descends the pending tree against SGLang's radix cache in one pass, returning all per-request hits in $O(C{\times}D)$ vs.\ a naive $O(N{\times}D)$; for high-sharing workloads $C \ll N$, an $N/C$ amortization\footnote{$N$ = queue size; $C$ = number of prefix-sharing clusters; $D$ = prompt depth in the tree ($\sim\!O(\log N)$).}. On vLLM's flat block-hash cache, PEEK falls back to per-request probes but still reads cluster signals off the tree---structure stock vLLM cannot see.

    \item \textbf{Cluster-aware admit and evict (\S\ref{sec:clpm_sched}).} PEEK's \emph{Cluster-LPM} (cLPM) extends LPM with pending-tree signals on top of its hit-length objective $h$. Requests in the same cluster therefore rank close together, so one prefill amortizes across the cluster before unrelated arrivals can evict the prefix---instead of scattering by arrival order and thrashing the cache (e.g., stock LPM on hit-length ties; \S\ref{sec:thrashing}). The eviction hook reads the same tree from the other side: zero-demand blocks evict first; blocks the largest pending clusters depend on are protected last.

    \item \textbf{Bound starvation, adaptively (\S\ref{sec:clpm_sched}).} We observed that cluster-first ordering can starve singletons. PEEK adds a fairness lane keyed by arrival, stride-interleaved with cLPM, and a \emph{dynamic-lane} controller widens the fairness share as singletons accumulate or as the oldest singleton's wait nears the SLO (EMA-smoothed). Thus, cluster locality dominates on prefix-rich queues; fairness ramps up only when the queue tilts singleton-heavy.
\end{enumerate}

\begin{table}[!t]
\centering
\small
\setlength{\tabcolsep}{1.5pt}
\renewcommand{\arraystretch}{0.9}
\begin{tabular}{lcccc}
\toprule
 & \textbf{FCFS} & \textbf{LPM} & \textbf{DFS\_W} & \textbf{PEEK} \\
 & (vLLM/SGLang) & (SGLang) & (SGLang) & (Ours) \\
\midrule
\multicolumn{5}{l}{\emph{Scheduling}} \\
\quad Prefix-aware & \ding{55} & Cache tree\textsuperscript{$\ddagger$} & Cache tree\textsuperscript{$\ddagger$} & Queue trie \\
\quad Ordering & Arrival & Per-request & Batch & Global \\
\midrule
\multicolumn{5}{l}{\emph{Caching \& Eviction}} \\
\quad Mechanism & APC / RadixCache & RadixCache & RadixCache & Pending trie \\
\quad Eviction & LRU/LFU & LRU/LFU & LRU/LFU & Queue-aware \\
\bottomrule
\end{tabular}
\caption[Scheduling and eviction mechanisms in current serving engines]{Scheduling and eviction mechanisms in current serving engines. PEEK is the only engine that jointly optimizes scheduling and eviction. {\scriptsize \textsuperscript{$\ddagger$}LPM/DFS\_W see only cache state, not queue sharing.}}
\label{tab:policy-comparison}
\end{table}

We evaluate PEEK on five workloads spanning single-GPU and multi-GPU deployments (up to 4$\times$H100, DP=2 over TP=2) across SGLang and vLLM (Appendix Table~\ref{tab:per_workload}). Our key findings:
\begin{itemize}[leftmargin=*,itemsep=0pt,topsep=2pt,parsep=0pt]
    \item \textbf{Substantial wins where prefix structure exists (\S\ref{sec:w1}--\S\ref{sec:w3}).} Across shared-prompt chat (W1), long-prefix RAG (W2), and multi-GPU 70B variants of W1 and W2 (W3; 4$\times$H100, DP=2$\times$TP=2), PEEK lifts cache hit \textbf{3.0$\times$/2.6$\times$}, TTFT \textbf{7.9$\times$/7.1$\times$}, E2E \textbf{6.7$\times$/5.5$\times$}, and throughput \textbf{3.6$\times$/4.5$\times$} (SGLang/vLLM); gains scale with KV pressure (2--16$\times$) and parallelism.
    \item \textbf{No regression on low-headroom traffic (\S\ref{sec:noregress}).} On agentic conversation bursts already prefix-coherent by arrival (W4) and singleton-dominated chat (W5), PEEK matches the strongest stock baselines within noise on E2E and throughput (\textbf{$\pm$2\%} on W4, \textbf{$\pm$2\%} on W5; TTFT envelopes are wider, $\pm$5--7\%) and modestly wins where weak overlap exists (W5 cell\_C\_long: $+$10\,pp cache hit, up to $\mathbf{2.6\times}$ TTFT on vLLM)---never regressing.
    \item \textbf{Cluster-aware admission is the dominant lever; other components refine it (\S\ref{sec:w1}).} cLPM alone lands within \textbf{3\,pp} cache hit of the full stack. Group-major (GM) shifts cache hit by \textbf{$-$0.1 to $+$1.4\,pp} (SGLang) / \textbf{$-$3.0 to $-$0.4\,pp} (vLLM) but costs \textbf{10--29\%} TTFT/E2E from singleton deferral; dynamic-lane (DL) fairness recovers \textbf{2.4--15.1\%}; queue-aware eviction (PE) further trims TTFT/E2E by \textbf{0--3\%} on SGLang, flat on vLLM.
\end{itemize}

PEEK also supports the \emph{offline} (batch) regime, resting on the same observation: the pending queue makes future cache demand \emph{predictive}, and the prefix relationships among waiting requests are the highest-signal structure to schedule and evict against. In contrast, PEEK-offline builds a one-shot prefix trie over the full submitted batch, DFS-traverses it to place sharing requests adjacently from a cold cache, and refines wave-level ordering against a shadow LRU model with the same queue-aware eviction hooks (Appendix~\ref{app:offline_method}). PEEK-offline delivers up to \textbf{1.6$\times$/1.8$\times$} throughput on single-GPU (Qwen2.5-14B/1$\times$H100) and \textbf{2.4$\times$/8.0$\times$} on multi-GPU (Qwen2.5-72B/2$\times$H100) over the strongest baselines (SGLang/vLLM), with near-zero overhead on non-sharing workloads (Appendix~\ref{app:offline_experiments}).

%% file: content/background.tex
\section{Background and Motivation}
\label{sec:background}
\vspace{-0.3em}

\vspace{-0.5em}
\paragraph{Prefix caching mechanisms.}
We focus on SGLang and vLLM, the leading open-source LLM serving engines in production use; a \emph{request} is a single completion call (prefill over the input prompt followed by autoregressive decoding). Both engines use \emph{PagedAttention} to decouple logical token sequences from physical KV cache blocks. On top of this, vLLM employs \textbf{APC} (Automatic Prefix Caching), which hashes each block by token content and prefix position via chained hashing to identify reusable blocks. SGLang adopts \textbf{RadixCache}, organizing cached blocks in a radix tree for $\mathcal{O}(D)$ per-request LPM. These mechanisms provide the \emph{architectural possibility} for prefix sharing; actual performance depends on the \emph{scheduling} and \emph{eviction} policies discussed below.

\vspace{-0.5em}
\paragraph{Scheduling and eviction policies.}
vLLM uses \textbf{FCFS} (arrival order, $\mathcal{O}(1)$, prefix-agnostic); SGLang offers \textbf{LPM} (Longest Prefix Match, $\mathcal{O}(N \cdot D)$ per round) and \textbf{DFS\_WEIGHT}, which groups requests by cache-tree position. All three bind to the \emph{current} cache state and ignore inter-queue sharing: FCFS+APC barely helps under cache pressure (\S\ref{sec:w1}); LPM (greedy per-round) and DFS\_WEIGHT (cache-tree-position grouping) both decide from current cache state and miss waiting-queue sharing. On the eviction side, both engines default to \textbf{LRU}, designed for OS page caches with unknown future access---but the LLM waiting queue encodes upcoming work, and LRU may evict prefixes that pending requests need next. SGLang's \textbf{LFU} alternative is similarly queue-oblivious. Under heavy contention, all of these struggle (\S\ref{sec:w1}).

\vspace{-0.5em}
\paragraph{Scheduling-induced thrashing.}
\label{sec:thrashing} Stock LPM ranks by hit length $h$ and breaks ties by arrival---reactive, blind to future queue demand. Consider 8 cold chat requests interleaved across three tenants ($A_1, B_1, C_1, A_2, B_2, C_2, A_3, B_3$; cluster sizes 3/3/2; each tenant's $\sim$2K-token prompt uncached) on a KV cache that holds only one prompt at a time. All tied at $h{=}0$, LPM admits in arrival order: prefill $A$, evict for $B$, evict for $C$, refill $A$---\textbf{0\% hit}. PEEK reorders the queue by cluster via the pending tree to $A_1, A_2, A_3, B_1, B_2, B_3, C_1, C_2$: each prompt prefills once and is reused---\textbf{62.5\% hit}. At 100+ tenants and $2$--$16\times$ KV pressure, the gap widens: cache-state-bound schedulers cannot anticipate which prefix the queue will demand next (\S\ref{sec:w1}).

\noindent\textbf{The gaps.}\quad \emph{(i) Visibility:} existing schedulers ignore inter-queue sharing. \emph{(ii) Coupling:} scheduling and eviction are decoupled. \emph{(iii) Co-design:} scheduling and eviction share the same queue-demand signal but no engine exploits it jointly. PEEK closes all three: an incremental radix tree over the waiting queue exposes prefix-sharing clusters the engines cannot see, driving cluster-aware admission and a co-designed eviction hook that protects blocks queued requests will reuse (\S\ref{sec:method}).

%% file: content/method.tex
\section{PEEK: \underline{P}r\underline{e}dictiv\underline{e} Queue-Informed \underline{K}V Cache Management}
\label{sec:method}

\begin{figure}[t]
\centering
\begin{tikzpicture}[
    node distance=0.45cm and 0.6cm,
    box/.style={draw, rounded corners, minimum height=0.7cm, align=center, font=\footnotesize},
    component/.style={box, fill=orange!10, draw=orange!50, thick, minimum width=2.6cm},
    io/.style={box, fill=blue!6, draw=blue!40, minimum width=2.4cm},
    engine/.style={box, fill=green!8, draw=green!50!black, thick, minimum width=2.6cm},
    cache/.style={box, fill=red!6, draw=red!40, minimum width=2.6cm, align=center},
    region/.style={draw, dashed, rounded corners=4pt, inner sep=6pt},
    arrow/.style={-{Stealth[length=2mm]}, thick},
    ann/.style={font=\scriptsize\bfseries, text=gray!40!black},
]
    \node[io] (queue) {Waiting Queue\\[-1pt]{\scriptsize $\{r_1, \ldots, r_N\}$}};
    \node[component, below=0.7cm of queue, xshift=-2.2cm] (tree)
        {Pending Radix Tree\\[-1pt]{\scriptsize incremental, $O(D)$/op (\S\ref{sec:pending_tree})}};
    \node[cache, below=0.7cm of queue, xshift=2.2cm] (cache)
        {Engine Prefix Cache\\[-1pt]{\scriptsize SGLang radix / vLLM hash}};
    \draw[arrow, dashed, gray!60!black] (tree.east) -- node[ann, above] {dual-walk (\S\ref{sec:dual_walk})} (cache.west);
    \draw[arrow, dashed, gray!60!black] (cache.west) -- (tree.east);
    \node[component, below=0.9cm of tree] (sched)
        {Cluster-Aware Scheduler\\[-1pt]{\scriptsize cLPM + multi-lane (\S\ref{sec:clpm_sched})}};
    \node[component, below=0.9cm of cache, fill=orange!10] (evict)
        {Eviction Hook\\[-1pt]{\scriptsize demand $\times$ depth (\S\ref{sec:eviction})}};
    \draw[arrow] (tree.south) -- (sched.north);
    \draw[arrow] (tree.south) -- ++(0, -0.45) -| (evict.north);
    \draw[arrow] (cache.south) -- ++(0, -0.45) -| (sched.north);
    \node[engine, below=0.9cm of sched, xshift=2.2cm] (engine)
        {Serving Engine\\[-1pt]{\tiny SGLang / vLLM}};
    \draw[arrow] (sched.south) -| (engine.west);
    \draw[arrow, orange!70!black, dashed] (evict.south) -| node[ann, right, pos=0.3] {evict hint} (engine.east);
    \draw[region, orange!40] ([xshift=-0.4cm, yshift=0.25cm]tree.north west)
        rectangle ([xshift=0.4cm, yshift=-0.25cm]evict.south east);
    \node[font=\scriptsize\bfseries, orange!70!black, anchor=north west]
        at ([xshift=-0.35cm, yshift=0.2cm]tree.north west) {PEEK};
\end{tikzpicture}
\caption{PEEK architecture.}
\label{fig:peek-arch}
\end{figure}
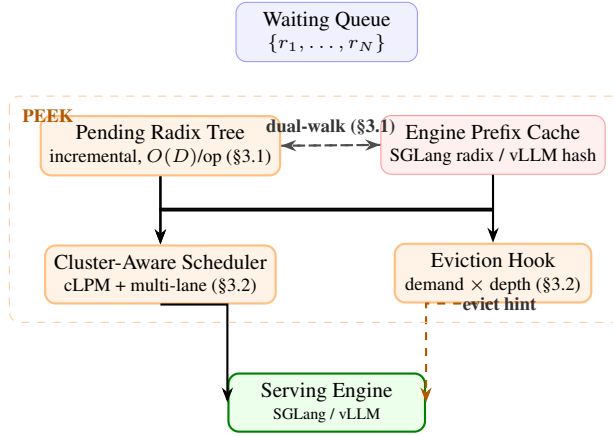

PEEK rests on a simple observation: \textbf{the waiting queue is itself a structured workload}. To utilize its signals, PEEK introduces an incremental radix tree over the pending queue (\S\ref{sec:pending_tree}) supporting three mechanisms (Figure~\ref{fig:peek-arch}): \textbf{(1)} a \emph{dual-walk} (\S\ref{sec:dual_walk}) yielding all-LPM via $O(C{\times}D)$ tree co-descent on SGLang's radix cache and per-request hash probes on vLLM's hash cache; \textbf{(2)} \emph{cluster-aware admission} (\S\ref{sec:clpm_sched}) admits the pioneer first so siblings reuse the cached prefix, paired with an \emph{eviction hook} that frees only blocks that no pending cluster needs; \textbf{(3)} a \emph{multi-lane stride scheduler} (\S\ref{sec:clpm_sched}) interleaving cache-locality with fairness.

\label{sec:impl}PEEK pairs a Rust core ($\sim$1.6\,kLOC; pending tree, dual-walk, cluster/score lookups; PyO3) with $\sim$800 LOC Python shims per engine that monkey-patch SGLang's \texttt{RadixCache}/\texttt{Scheduler} and vLLM v1's \texttt{Scheduler}/\texttt{BlockPool}; cLPM sort runs in Python over Rust-computed signals. Environment flags toggle PEEK per-binary, no upstream fork (Appendix~\ref{app:implementation}).

\subsection{Pending Radix Tree and Dual-Walk for LPM}
\label{sec:pending_tree}
\label{sec:dual_walk}

The pending radix tree is a compressed trie over the waiting prompts. Each leaf carries the request IDs (\emph{terminators}) whose prompt ends there; each internal node holds a \texttt{pending\_count}---the number of waiting prompts passing through. Edges store multi-token runs via radix compression, so size scales with distinct prefix branches, not raw token count. PEEK maintains the tree incrementally: \texttt{insert} on arrival, \texttt{remove} on admission/completion, both $O(D)$ along the prompt's path (typically $O(\log N)$). The tree is always current at scheduling time---no per-cycle rebuild. Two requests share a queued prefix iff their prompts pass through a common internal node with $\texttt{pending\_count} \geq 2$; the deepest such node is the request's \emph{cluster}, and its \texttt{pending\_count} the cluster size.

Each cycle, PEEK obtains the LPM hit $h(r)$ for every waiting request and separately reads its cluster signals (size, ancestor score) from the pending tree at $O(1)$ per request. On SGLang's radix cache, PEEK \emph{dual-walks}---tree-vs-tree co-descent of the pending tree against the radix cache (Algorithm~\ref{alg:dual_walk}; Figure~\ref{fig:dualwalk} in Appendix~\ref{app:dual_walk_alg})---collapsing $N$ per-request lookups into $C$ per-cluster lookups, $O(C{\times}D)$ vs.\ $O(N{\times}D)$, an amortization of $N/C$ that is large for high-sharing workloads where $C \ll N$. On vLLM's flat block-hash cache, no structural co-descent is possible: PEEK calls the engine's per-request \texttt{find\_longest\_cache\_hit} once per request ($O(N{\times}b)$, $b$ = block-hash chain length), then reads cluster signals separately from the pending tree---an $O(N)$ traversal exposing the cluster structure stock vLLM cannot see.

\begin{algorithm}[t]
\caption{Dual-Walk on SGLang: per-cycle LPM via tree-vs-tree co-descent}
\label{alg:dual_walk}
\begin{algorithmic}[1]
\REQUIRE Pending tree $T$; SGLang radix cache $\mathcal{C}$
\ENSURE LPM hit length $H[r]$ for every waiting request $r$
\STATE $H \leftarrow \{\}$;\quad \textsc{Visit}($T.\text{root}$, $0$);\quad \textbf{return} $H$
\STATE \textbf{procedure} \textsc{Visit}(pending node $u$, matched depth $d$):
\FOR{each child edge $e = (u \to u')$ of $u$}
    \STATE $\delta \leftarrow \textsc{MatchInRadix}(e.\text{tokens},\ \mathcal{C},\ d)$ \COMMENT{co-traverse $\mathcal{C}$ from depth $d$}
    \STATE $h \leftarrow d + \delta$
    \FOR{each terminator $r$ in $\textsc{Subtree}(u')$}
        \STATE $H[r] \leftarrow \max(H[r],\ h)$ \COMMENT{LPM along this branch}
    \ENDFOR
    \IF{$\delta = |e.\text{tokens}|$} \STATE \textsc{Visit}($u'$, $h$) \COMMENT{full edge matched: descend} \ENDIF
\ENDFOR
\STATE \textbf{vLLM variant.} The flat block-hash cache permits no co-descent. Instead, PEEK calls the engine's per-request \texttt{find\_longest\_cache\_hit} for each $r$ to obtain $H[r]$ in $O(N{\times}b)$ ($b$=block-hash chain length), then walks the pending tree separately ($O(N)$) to read cluster signals.
\end{algorithmic}
\end{algorithm}

\subsection{Pending-Tree-Driven Scheduling and Eviction}
\label{sec:clpm_sched}
\label{sec:eviction}

PEEK's scheduler (cLPM) and eviction hook (PE) both read the same pending tree, so admission and protection are consistent without explicit coordination. The full per-cycle procedure is given in Algorithm~\ref{alg:clpm} (Appendix~\ref{app:clpm_algorithm}); we walk through its components below.

For every waiting request $r$, cLPM produces three signals: (1) the \emph{LPM hit} $h(r)$ against the engine cache; (2) the \emph{section} $r$ belongs to---\emph{warm} if $h(r){>}\tau$, otherwise \emph{pioneer} if $r$'s first $\theta{=}32$ tokens have not yet been claimed this cycle, else \emph{sibling}; (3) two pending-tree statistics, $|\texttt{cluster}(r)|$ (pending count at $r$'s deepest $\geq$2 ancestor) and the \emph{ancestor score} $\texttt{score}(r){=}\sum_{v \in \pi(r)} \texttt{pending\_count}(v)\!\cdot\!|\texttt{edge}(v)|$ (depth-weighted demand along $r$'s path). cLPM then sorts by
\begin{equation}
\label{eq:clpm_key_a}
    \texttt{key}(r) = \big(\,\texttt{section}(r),\; -h(r),\; -\texttt{score}(r),\; -|\texttt{cluster}(r)|,\; \texttt{arrival}(r)\,\big).
\end{equation}

By comparison, stock LPM is a special case: it sorts by $(-h(r),\, \texttt{is\_depr}(r),\, \texttt{arrival}(r))$, where $\texttt{is\_depr}$ is the same $\theta$-prefix pioneer/sibling flag. cLPM differs in two ways: it \emph{promotes} the section distinction from a binary tiebreak to the primary key (so any pioneer admits before any sibling, regardless of $h$), and it \emph{adds} the cluster-aware tiebreaks $-\texttt{score}$ and $-|\texttt{cluster}|$ (which LPM cannot compute---it has no pending tree).

Against the thrashing failure mode (\S\ref{sec:thrashing}), cLPM's score ordering keeps a larger cluster's members together \emph{within} their section, amortizing each prefill across multiple hits---making cLPM the dominant cache-hit lever (within 3\,pp of the full PEEK stack, \S\ref{sec:w1}). PEEK additionally supports an optional strengthening, \emph{group-major} (GM), that emits all members of one cluster contiguously (1 prefill $+$ $\texttt{size}{-}1$ hits per cluster); cLPM's score ordering already approximates GM grouping in typical queues, so GM adds only marginal cache hit.

We observed that GM helps cache hit but aggressively defers singletons. To rescue them, PEEK introduces a \emph{fairness lane} and a \emph{dynamic-lane} (DL) controller: cLPM stride-interleaves Lane $A$ (Eq.~\ref{eq:clpm_key_a}) with the fairness lane keyed by $(\texttt{section},\, \texttt{arrival},\, -h)$: a fraction $\alpha$ of admissions come from Lane $A$, $1{-}\alpha$ from Lane $B$. The oldest waiting request advances at least every $\lceil 1/(1{-}\alpha)\rceil$ cycles. Default $\alpha{=}0.7$. DL adjusts $\alpha$ per cycle from queue composition---\texttt{singleton\_frac} (fraction in no cluster) and \texttt{age\_pressure}${=}\min(1,\,\texttt{oldest\_singleton\_wait}/2\text{s})$---ramping the fairness share $1{-}\alpha$ from 0.15 (cluster-dominated) toward 0.60 (singleton-heavy or near-SLO), EMA-smoothed:
\begin{equation*}
\!\!\!1{-}\alpha \leftarrow \texttt{EMA}_{\beta=0.3}\!\big(\texttt{clamp}(0.15 + 0.5\!\cdot\!\texttt{singleton\_frac} + 0.3\!\cdot\!\texttt{age\_pressure},\,0.10,\,0.60)\big).
\end{equation*}

For eviction, PEEK's hook (PE) overrides LRU/LFU when the engine must free memory, using the same pending tree: candidate block $b$ with tokens $t_b$ scores
\begin{equation}
\label{eq:evict_score}
    \texttt{evict\_score}(b) = \max_{v \in \texttt{ancestors}(t_b)} \texttt{pending\_count}(v) \cdot \texttt{depth}(v),
\end{equation}
depth-weighted demand along $b$'s ancestor chain (max captures deep lightly-shared prefixes sitting above heavily-shared subtrees); lowest score evicts first. Of four scoring modes (\emph{plain}, \emph{cluster}, \emph{recency}, \emph{decay}), experiments use \emph{cluster}. SGLang installs a custom \texttt{EvictionStrategy}; vLLM patches \texttt{BlockPool.get\_new\_blocks} to prefer zero-demand victims (capped at $4N$).

In summary, DL's clamp $[0.10, 0.60]$ keeps cache locality dominant (Lane $A$ retains at least 40\%) while EMA damps oscillation, rescuing singletons GM defers; PE protects what waiting clusters need by evicting zero-demand blocks first and keeping high-demand blocks last. Together, scheduling and eviction stay aligned through the same pending tree.

\subsection{Guards}
\label{sec:guards}

PEEK's per-cycle cost (dual-walk + sort) would otherwise apply even when the queue has no sharing. The pending tree exposes a \texttt{has\_sharing} primitive---an $O(\#\text{root-children})$ scan returning true iff some root-child has $\texttt{pending\_count} \geq 2$---available for callers to short-circuit cLPM. In addition, on SGLang the stock scheduler's queue-depth fallback (LPM degenerates to FCFS once the waiting queue exceeds 128 requests) provides a second cheap exit. Empirically these two paths together leave PEEK matching FCFS within $\pm$1.5\% on workloads with no exploitable structure (W4, W5; \S\ref{sec:noregress}); the dual-walk + sort cost is amortized to the noise floor in those regimes.

%% file: content/experimental_setup.tex
\section{Experiments}
\label{sec:experiments}

\label{sec:setup}

We evaluate PEEK on a five-workload suite (W1--W5) covering typical production LLM serving scenarios across diverse models, datasets, and single- and multi-GPU hardware (Appendix Tables~\ref{tab:hardware}--\ref{tab:per_workload},~\ref{tab:prod_mapping}). Each workload comprises one or more \emph{cells}---parameter configurations along its primary axis---each run at \emph{moderate} and \emph{heavy} load.\footnote{Moderate targets sustained queue length 60--127; heavy targets $\geq$128, the threshold at which SGLang's LPM falls back to FCFS. The same targets are used on vLLM for cross-engine comparability.} We report means across 3 seeds after $\sim$10--20\% warmup; Table~\ref{tab:setup_summary} lists policy labels. W1 (\S\ref{sec:w1}) sweeps the full ablation lattice and identifies \texttt{cLPM+GM+DL+PE} as PEEK's most performant policy; W2 (\S\ref{sec:w2}), W3 (\S\ref{sec:w3}), and W4--W5 (\S\ref{sec:noregress}) probe orthogonal axes using \texttt{cLPM+GM+DL+PE} alone, conserving GPU resources.

\begin{table}[t]
\centering
\footnotesize
\setlength{\tabcolsep}{4pt}
\begin{tabular}{lll}
\toprule
\multicolumn{3}{c}{\emph{Policy labels}} \\
\midrule
\textbf{Label} & \textbf{Components (scheduler + eviction)} & \textbf{Role} \\
\midrule
FCFS+LRU                      & FCFS + LRU                                & Na\"ive baseline (no prefix caching) \\
LPM+LRU                       & SGLang LPM + LRU                          & SGLang baseline \\
FCFS(APC)+LRU                 & vLLM FCFS + APC + LRU                     & vLLM baseline \\
LPM+PE / FCFS(APC)+PE         & stock scheduler + PEEK PE                 & Eviction-only ablation \\
cLPM                          & cLPM + LRU                                & Sort-key-only ablation \\
cLPM+GM                       & cLPM + group-major + LRU                  & Scheduling (sort + admission) \\
cLPM+GM+PE                    & cLPM + group-major + PEEK PE              & Primary co-design \\
cLPM+GM+DL+PE                 & cLPM + group-major + dynamic-lane + PE    & + starvation guard \\
\bottomrule
\end{tabular}
\caption{Policy labels used in Section~\ref{sec:experiments}.}
\label{tab:setup_summary}
\end{table}

%% file: content/results.tex
\subsection{W1: Shared-Prompt Chat Across Production KV-Pressure Tiers}
\label{sec:w1}

W1 tests whether PEEK's co-design holds up as KV-cache pressure rises. Cells A/B/C/D span 2$\times$ to 16$\times$ KV oversubscription, each mapped to a production class (Appendix Tables~\ref{tab:w1_cells}--\ref{tab:w1_cell_meaning}). Figures~\ref{fig:w1_cache_hit}--\ref{fig:w1_ttft} plot cache hit and TTFT; throughput and E2E are in Appendix Figures~\ref{fig:w1_rps},~\ref{fig:w1_e2e} and full numerical results in Appendix Tables~\ref{tab:w1_sglang}--\ref{tab:w1_decode}. We first use cLPM+GM+DL+PE to compare against the stock baselines.

\begin{figure}[!htb]
\centering
\begin{subfigure}[t]{0.48\columnwidth}\centering\includegraphics[width=\linewidth]{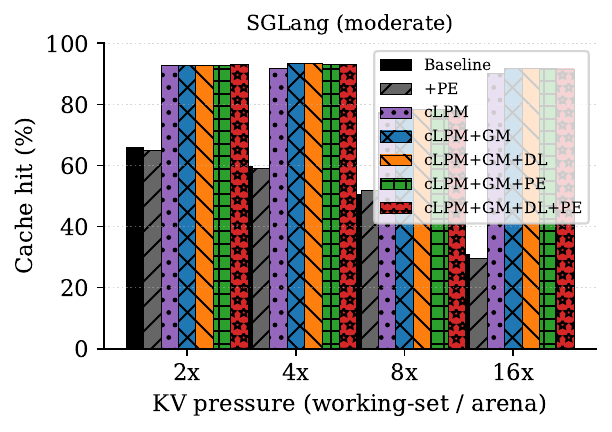}\caption{SGLang, moderate.}\label{fig:w1_sglang_cache_hit_mod}\end{subfigure}\hfill
\begin{subfigure}[t]{0.48\columnwidth}\centering\includegraphics[width=\linewidth]{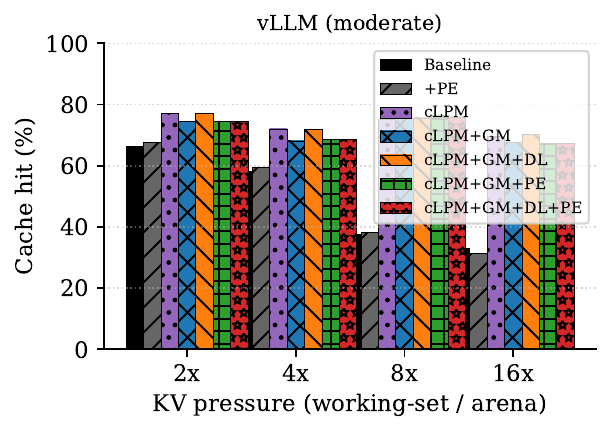}\caption{vLLM, moderate.}\label{fig:w1_vllm_cache_hit_mod}\end{subfigure}
\\[0.4em]
\begin{subfigure}[t]{0.48\columnwidth}\centering\includegraphics[width=\linewidth]{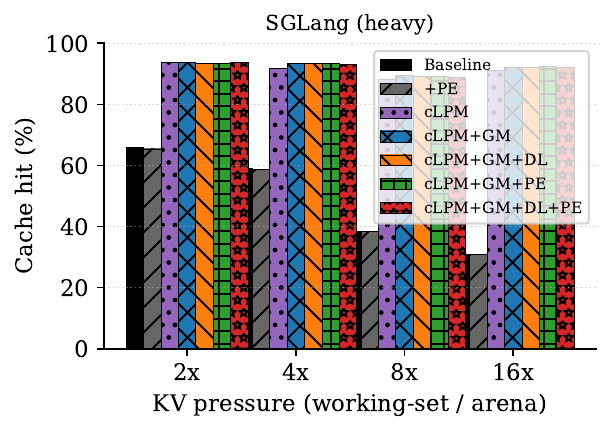}\caption{SGLang, heavy.}\label{fig:w1_sglang_cache_hit_heavy}\end{subfigure}\hfill
\begin{subfigure}[t]{0.48\columnwidth}\centering\includegraphics[width=\linewidth]{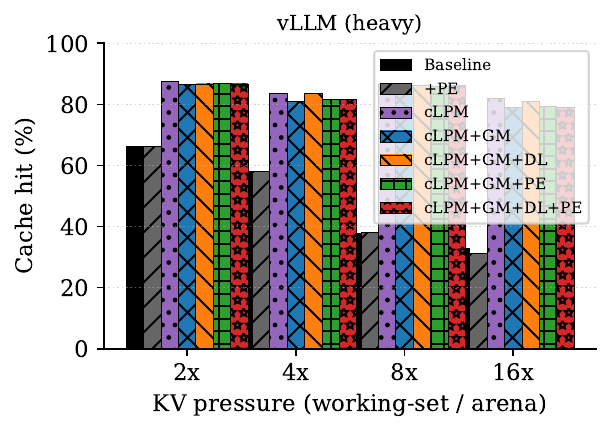}\caption{vLLM, heavy.}\label{fig:w1_vllm_cache_hit_heavy}\end{subfigure}
\caption{W1 cache hit rate vs.\ KV pressure.}
\label{fig:w1_cache_hit}
\end{figure}

\begin{figure}[!htb]
\centering
\begin{subfigure}[t]{0.48\columnwidth}\centering\includegraphics[width=\linewidth]{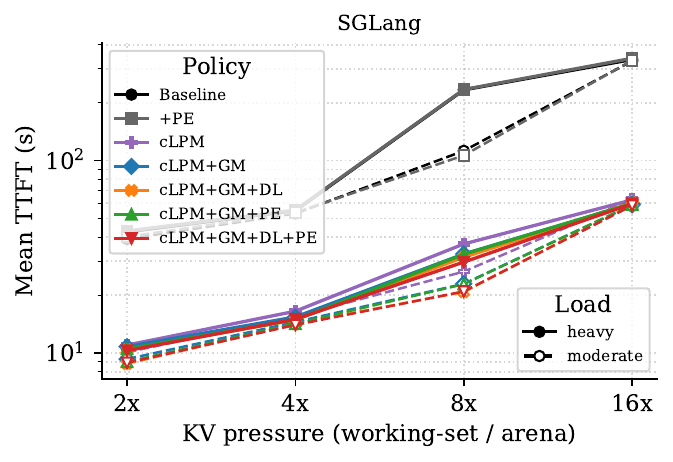}\caption{SGLang.}\label{fig:w1_sglang_ttft}\end{subfigure}\hfill
\begin{subfigure}[t]{0.48\columnwidth}\centering\includegraphics[width=\linewidth]{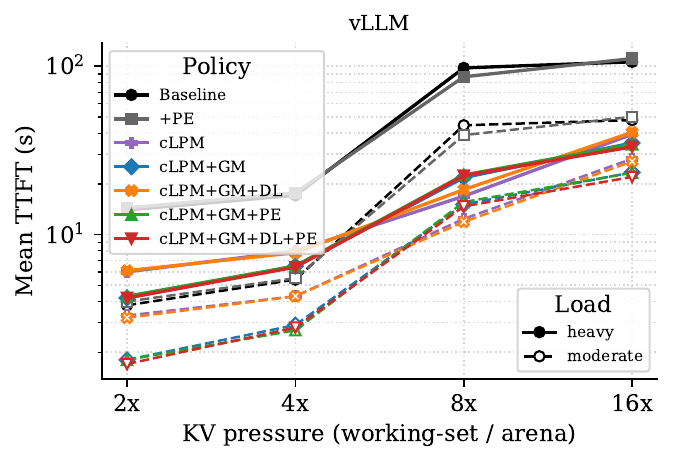}\caption{vLLM.}\label{fig:w1_vllm_ttft}\end{subfigure}
\caption{W1 mean TTFT vs.\ KV pressure (log y-axis). Each panel overlays both load levels: solid lines with filled markers = heavy, dashed lines with hollow markers = moderate.}
\label{fig:w1_ttft}
\end{figure}

cLPM+GM+DL+PE lifts cache hit by 28--61\,pp (SGLang, heavy) / 20--49\,pp (vLLM, heavy; moderate cells start as low as 8\,pp)---\textbf{1.4--3.0$\times$/1.3--2.4$\times$}---and these gains transfer to TTFT \textbf{3.7--7.9$\times$/1.9--4.4$\times$}, E2E \textbf{2.9--6.7$\times$/1.2--3.6$\times$}, and throughput \textbf{1.5--3.6$\times$/1.2--2.5$\times$} (SGLang/vLLM). Gains scale with KV pressure as baselines thrash (\S\ref{sec:thrashing}) while cLPM+GM+DL+PE stays flat, except at 16$\times$ heavy load where over-saturation narrows the ratio. Decode-side (Appendix Table~\ref{tab:w1_decode}): TPOT runs \textbf{8--26\%} (\textbf{4--12\,ms/token}) above baseline; ITL tracks TPOT closely (drift $+$3.6 to $+$11.5\,ms across cells) and moves with throughput rather than per-token cost---both costs are dwarfed by the E2E gains.

cLPM alone contributes most of PEEK's gain---within \textbf{3\,pp} of the full PEEK stack on cache hit on both engines---confirming that cluster-aware admission is the dominant lever among PEEK's mechanisms. The mechanism behind the gain: by replacing stock LPM's FCFS-on-ties with pending-tree-aware tiebreaking, cLPM admits cluster siblings adjacent to their pioneer, so each shared-prefix prefill is amortized across many cache hits before any unrelated arrival can evict the prefix.

cLPM+GM lifts SGLang cache hit \textbf{$-$0.1 to $+$1.4 pp} over cLPM but trails the best on TTFT/E2E by up to \textbf{10\%/8\%}; on vLLM it \emph{reduces} cache hit by \textbf{0.4--3.0 pp} and trails on TTFT/E2E by up to \textbf{29\%/22\%} at cell C. \textbf{GM, aggressive clustering, buys little cache hit but inflates TTFT/E2E}: cLPM already heads the queue with cluster pioneers, and GM locks in the entire cluster as a contiguous batch---amortizing the shared prefill across siblings while deferring singletons.

DL shortens TTFT/E2E by \textbf{2.4--15.1\%} on SGLang A/C and vLLM C, \textbf{recovering the singleton-starvation cost that GM imposes}; elsewhere it trades a small TTFT regression for queue-aging fairness. PE adds \textbf{0--3\%} on SGLang and is flat on vLLM ($\pm$5\%); on a stock scheduler PE is flat ($\pm$1\,pp, $\pm$2\%)---\textbf{eviction cannot create temporal locality the scheduler did not first establish}.
\vspace{-0.6em}
\paragraph{Takeaway 1.} For high-sharing workloads, all cLPM-based PEEK policies substantially outperform stock baselines on both engines on cache hit, TTFT, E2E, and throughput, with the gain scaling as KV pressure rises. The biggest gain comes from cLPM's cluster-aware scoring atop stock LPM. GM further pushes cache hit but at the cost of delaying singletons, causing a noticeable loss on TTFT and E2E. DL and PE recover those two ends, ultimately establishing cLPM+GM+DL+PE as the most performant policy on both engines.

\subsection{W2: Long-Document RAG under Decode-Dominant KV Pressure}
\label{sec:w2}

W2 tests PEEK on long-prefix RAG, where documents overflow the KV arena (Appendix Table~\ref{tab:per_workload}). Figure~\ref{fig:w2_cache_hit_ttft} plots cache hit and TTFT; throughput and E2E are in Appendix Figures~\ref{fig:w2_rps},~\ref{fig:w2_e2e} and full numerical results in Appendix Tables~\ref{tab:w2_sglang}--\ref{tab:w2_vllm}.

\begin{figure}[!htb]
\centering
\begin{subfigure}[t]{0.48\columnwidth}\centering\includegraphics[width=\linewidth]{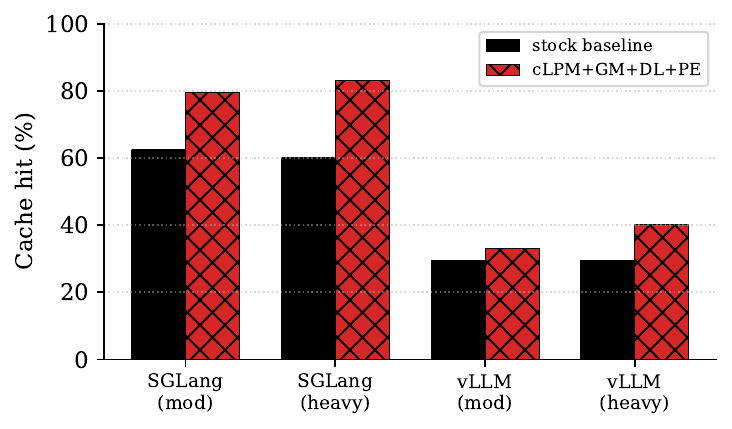}\caption{Cache hit (SGLang + vLLM).}\label{fig:w2_cache_hit}\end{subfigure}\hfill
\begin{subfigure}[t]{0.48\columnwidth}\centering\includegraphics[width=\linewidth]{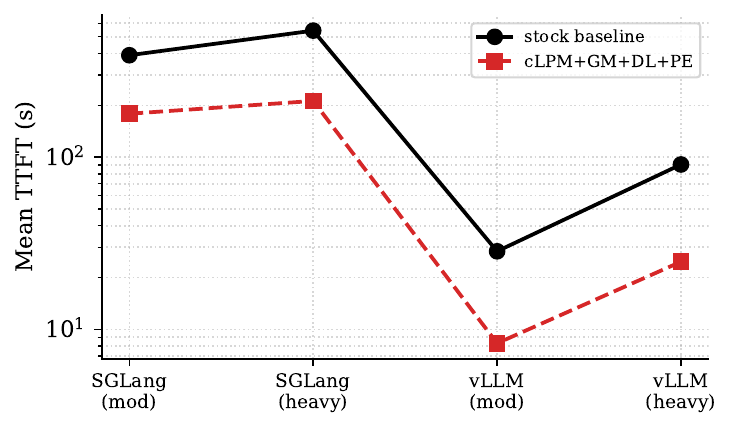}\caption{Mean TTFT (SGLang + vLLM, log y-axis).}\label{fig:w2_ttft}\end{subfigure}
\caption{W2 cache hit and mean TTFT, both engines and load levels.}
\label{fig:w2_cache_hit_ttft}
\end{figure}
\vspace{-0.6em}

cLPM+GM+DL+PE substantially outperforms the per-engine baseline on long-prefix RAG. Cache hit lifts \textbf{$+$17/$+$23\,pp} (SGLang, mod/heavy) and \textbf{$+$3.6/$+$10.6\,pp} (vLLM)---equivalently \textbf{1.27$\times$/1.39$\times$} and \textbf{1.12$\times$/1.36$\times$}. These translate into TTFT reductions of \textbf{2.2$\times$/2.6$\times$} on SGLang and \textbf{3.4$\times$/3.7$\times$} on vLLM, mean E2E reductions of \textbf{2.0$\times$/2.4$\times$} and \textbf{1.6$\times$/2.3$\times$}, and throughput rising \textbf{1.14$\times$/1.26$\times$} on SGLang and \textbf{1.00$\times$/1.06$\times$} on vLLM. The large TTFT and E2E reduction (\textbf{2$\times$--3.7$\times$}) on both engines reflects the same pattern as W1: under thrashing, low cache hit both inflates per-request prefill and blocks the queue, so cLPM+GM+DL+PE accelerates admitted requests \emph{and} clears the queue faster (\S\ref{sec:thrashing}). Decode-side, mean TPOT moves within \textbf{$\pm$5\%} on SGLang and \textbf{$\pm$9\%} on vLLM; mean ITL agrees with TPOT to within \textbf{1\,ms} in every cell---the gain comes from prefill recovery rather than decode-side optimization.

\vspace{-0.6em}
\paragraph{Takeaway 2.} As in W1, cLPM+GM+DL+PE wins large on both engines, confirming \textbf{PEEK's advantage transfers to long-decode RAG scenarios}.

\subsection{W3: Multi-Cell Shared-Prompt Workloads (Multi-GPU, DP=1 and DP=2)}
\label{sec:w3}

\begin{figure}[!htb]
\centering
\begin{subfigure}[t]{0.48\columnwidth}\centering\includegraphics[width=\linewidth]{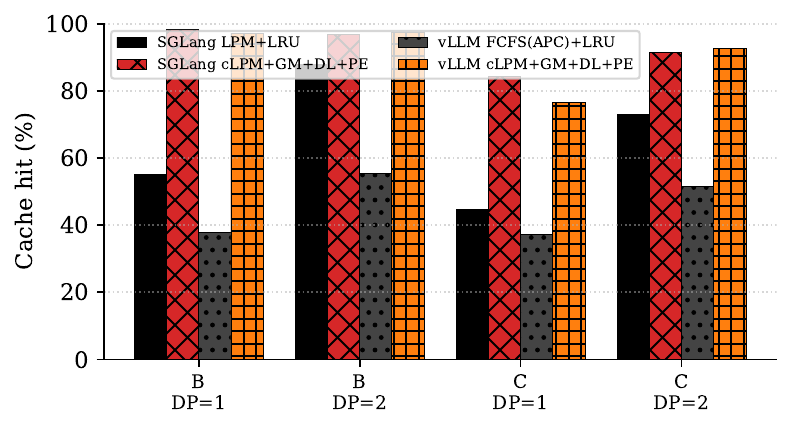}\caption{Cache hit (SGLang + vLLM).}\label{fig:w3_cache_hit}\end{subfigure}\hfill
\begin{subfigure}[t]{0.48\columnwidth}\centering\includegraphics[width=\linewidth]{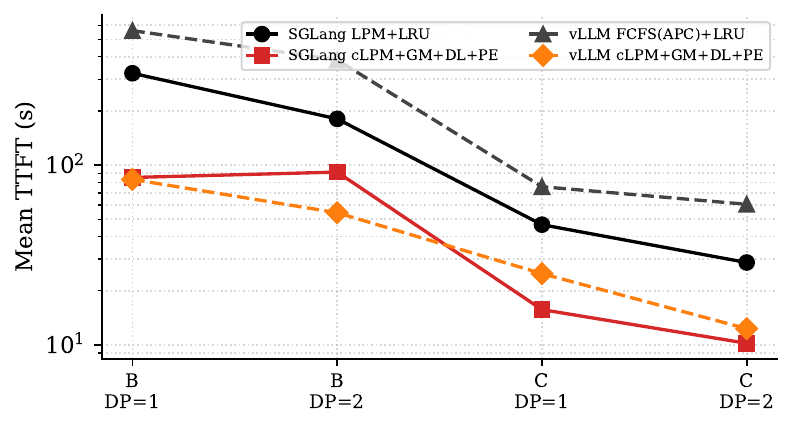}\caption{Mean TTFT (SGLang + vLLM, log y-axis).}\label{fig:w3_ttft}\end{subfigure}
\caption{W3 cache hit and mean TTFT across two cells (B, C) and two topologies (DP=1 / 2$\times$H100, DP=2 / 4$\times$H100), heavy load.}
\label{fig:w3_cache_hit_ttft}
\end{figure}

W3 scales W1 and W2 to Llama-3.1-70B-Instruct (TP=2) at two data-parallel scales---DP=1 (single replica, 2$\times$H100) and DP=2 (two replicas, 4$\times$H100)---across two regimes: \emph{cell B} mirrors W2 (RAG-like, decode-bound) and \emph{cell C} mirrors W1 (chat-like, admission-bound); see Figure~\ref{fig:w3_cache_hit_ttft}, Appendix Figures~\ref{fig:w3_rps}--\ref{fig:w3_e2e} (\S\ref{app:w3_figs}), Appendix Tables~\ref{tab:w3_sglang}--\ref{tab:w3_vllm} (\S\ref{app:w3_tables}), and Appendix Table~\ref{tab:per_workload}.

\vspace{-0.6em}

cLPM+GM+DL+PE wins large at \emph{both} topologies. At DP=1, SGLang cache hit lifts \textbf{$+$43.1/$+$39.6\,pp} on cell B/C, TTFT \textbf{3.80$\times$/2.97$\times$}, E2E \textbf{3.62$\times$/2.65$\times$}, throughput \textbf{2.66$\times$/2.08$\times$}; vLLM cache hit lifts \textbf{$+$59.4/$+$39.3\,pp}, TTFT \textbf{6.73$\times$/3.04$\times$}, E2E \textbf{5.54$\times$/2.58$\times$}, throughput \textbf{4.45$\times$/2.07$\times$}. Doubling to DP=2 mostly helps the stock baseline: SGLang baseline cache hit jumps \textbf{$+$32.7/$+$28.3\,pp} from DP=1 to DP=2 (vLLM \textbf{$+$17.7/$+$14.2\,pp}), while PEEK is already at ceiling on cell B (96.9--98.2\% SGLang, 97.1--97.3\% vLLM) and reaches 84--92\% on cell C. The relative gap thus narrows on SGLang to \textbf{$+$9.1/$+$18.6\,pp} (still up to 2.81$\times$ TTFT, 1.62--1.76$\times$ throughput) but stays nearly flat on vLLM (\textbf{$+$41.9/$+$41.3\,pp}; up to \textbf{7.08$\times$} TTFT, \textbf{4.50$\times$/2.83$\times$} throughput)---APC's block-hash misalignment doesn't shrink with KV space. Notably, PEEK \emph{at DP=1} matches or beats the stock baseline \emph{at DP=2} on cell B for both engines (SGLang: 98.2 vs 87.8\% cache, 85 vs 181\,s TTFT, 1.24 vs 1.04 req/s; vLLM: 97.1 vs 55.4\%, 83 vs 384\,s, 1.07 vs 0.48 req/s)---PEEK delivers more on half the GPUs than the per-engine baseline does on four. Decode-side, TPOT moves within \textbf{$\pm$7\%} on SGLang (cell C/DP=1 the worst at $-$6.9\%); on vLLM TPOT rises with throughput on PEEK runs (larger decode batch inflates per-token time in absolute terms, but prefill dominates and E2E still drops sharply).

\vspace{-0.6em}
\paragraph{Takeaway 3.} cLPM+GM+DL+PE wins large at both DP=1 and DP=2 on Llama-3.1-70B. \textbf{PEEK's gain scales to larger models and multi-GPU topologies}.

\subsection{No-Regress on Prefix-Coherent and Singleton-Dominated Traffic (W4, W5)}
\label{sec:noregress}

W4 and W5 establish PEEK's no-regress claim from two angles. \textbf{W4} runs an already-prefix-coherent scenario (\emph{agentic\_only}: multi-turn bursts; \emph{agentic\_shared}: 1402-token shared system prompt), where session-burst adjacency leaves little room to improve. \textbf{W5} runs heterogeneous singleton chat in \emph{cell\_C\_long} (512--4096 tokens, weak residual overlap) and \emph{cell\_C\_short} (32--1024 tokens, no structural sharing). Full results in Appendix~\S\ref{app:w4_tables}--\ref{app:w4_figs} (W4) and \S\ref{app:w5_figs}--\ref{app:w5_tables} (W5).

\textbf{W4 (already prefix-coherent).} cLPM+GM+DL+PE matches or marginally beats the strongest stock baseline (LPM+LRU on SGLang, FCFS(APC)+LRU on vLLM) on every metric in every cell: SGLang cache hit \textbf{$+$0.3--1.0\,pp}, TTFT within \textbf{$\pm$5\%}, E2E \textbf{$-$0.8--2.5\%}, throughput \textbf{$+$1.1\%}; vLLM cache hit \textbf{$+$0.0--1.7\,pp}, TTFT within \textbf{$\pm$7\%}, E2E and throughput within \textbf{$\pm$2\%}. TPOT moves \textbf{$-$0.3--3.5\%} (SGLang) / \textbf{$\pm$2\%} (vLLM). Marginal wins land where arrival order mismatches prefix order: SGLang on \emph{agentic\_shared}, vLLM on \emph{agentic\_only} (APC hash collisions).

\textbf{W5 (singleton-dominated).} On \emph{cell\_C\_long}, cluster-aware sort detects weak overlap LPM's exact match misses: SGLang cache hit \textbf{$+$9.5\,pp} (0.7\%$\rightarrow$10.2\%), TTFT \textbf{$-$11\%}, E2E \textbf{$-$8\%}, throughput \textbf{$+$5\%}; vLLM TTFT \textbf{2.6$\times$} reduction, E2E \textbf{$-$20\%}, throughput \textbf{$+$21\%} (APC reuse 97.5\%$\rightarrow$98.8\%). On \emph{cell\_C\_short}, the \texttt{has\_sharing} guard (\S\ref{sec:guards}) short-circuits PEEK to FCFS: every metric within \textbf{$\pm$2\%} except SGLang TTFT $+1.7\%$ and vLLM TTFT $-10\%$ relative ($0.04$\,s absolute).

\vspace{-0.6em}
\paragraph{Takeaway 4.} PEEK is no-regress on W4 and W5: the \texttt{has\_sharing} guard short-circuits to FCFS when no structure exists, while cluster-aware admission captures residual overlap when it does. Net: within \textbf{$\pm$2\%} (W4) and \textbf{$\pm$1\%} (W5) of the strongest baseline, with modest wins on weak structure---complementing W1--W3, where PEEK \emph{recovers} locality the baseline destroyed.

%% file: content/related_work.tex
\section{Related Work}
\label{sec:related_work}
\vspace{-0.5em}

\noindent\textit{Prefix-aware request scheduling.}
\emph{Online:} Preble~\citep{srivatsa2024preble} co-optimizes KV reuse with distributed load balancing across replicas. \citet{fu2024efficient} reorder by predicted output length to reduce head-of-line blocking. DLPM~\citep{cao2025dlpm} adds per-client deficit counters to LPM, bounding the service difference between clients while preserving prefix locality. \citet{dexter2025scheduling} prove the prefix-aware scheduling problem NP-hard and propose a $k$-LPM approximation. HotPrefix~\citep{li2025hotprefix} drives admission and CPU/GPU promotion from a moving average of past hits (\emph{cache hotness}). SOLA~\citep{hong2025sola} arbitrates prefix grouping against per-request SLOs. Autellix~\citep{luo2025autellix} reorders at the program level for agent workloads, capturing inter-call dependencies. \emph{Offline / batch:} BatchLLM~\citep{zheng2024batchllm} groups requests by global prefix sharing for high-throughput batched inference. BlendServe~\citep{zhao2026blendserve} reorders an offline workload to jointly maximize prefix/KV reuse and balance compute- vs. memory-bound phases, exploiting full visibility into the request set before submission. PEEK differs by forming groups from the \emph{pending queue itself} via an incremental prefix trie---robust to cold, stale, or contended cache---and uses the same trie for eviction so the two co-optimize one signal; it is intra-replica and composes beneath cross-replica routers, fairness contracts, or SLO arbiters. PEEK-offline (Appendix~\ref{app:peek_offline}) sits in the BatchLLM/BlendServe regime; this paper targets the online case with no full-workload visibility.

\noindent\textit{KV cache reuse across prompts.}
A separate axis supplies the \emph{capability} to share KV across prompts: PromptCache~\citep{gim2024prompt} (module-level KV via positional re-encoding), RAGCache~\citep{jin2024ragcache} (hierarchical RAG cache), ChunkAttention~\citep{ye2024chunkattention} (attention over a prefix tree), Hydragen~\citep{juravsky2024hydragen} (fused shared-prefix kernel), CachedAttention~\citep{gao2024cachedattention} (session-scoped state), CacheBlend~\citep{yao2024cacheblend} (block reuse across retrieved chunks), CacheGen~\citep{liu2024cachegen} (compressed KV streaming), LMCache~\citep{lmcache2025} (distributed shared KV). These are mechanisms for \emph{how} KV is stored; PEEK is the policy for \emph{which} blocks are cached and in what order requests run, composing atop any of them.

\noindent\textit{Within-sequence KV compression and eviction.}
H2O~\citep{zhang2024h2o} keeps heavy-hitter tokens by attention history; SnapKV~\citep{li2024snapkv} clusters and prunes tokens at prefill; Scissorhands~\citep{liu2024scissorhands} drops decayed-attention tokens; InfiniGen~\citep{lee2024infinigen} offloads cold KV with prefetch; KIVI~\citep{yuan2024kivi} quantizes KV to 2-bit. These act per-request along the sequence dimension; PEEK acts across requests along the block-allocation dimension and is fully orthogonal.

\noindent\textit{Disaggregated and distributed serving.}
DistServe~\citep{zhong2024distserve} and Splitwise~\citep{patel2024splitwise} disaggregate prefill from decode; Mooncake~\citep{qin2024mooncake} extends this to a distributed KV pool; Llumnix~\citep{sun2024llumnix} migrates running requests across instances. These reshape \emph{where} requests run; PEEK reshapes \emph{when} they run within a replica, and the two axes compose.

\noindent\textit{Cross-replica prefix-affinity routing.}
Gateways such as llm-d~\citep{llmd2025} score backends by prefix-affinity, queue depth, and KV utilization to route requests to the replica most likely to hit warm cache. This is cross-replica; PEEK improves intra-replica scheduling and sits beneath such a router.

\noindent\textit{Future-aware cache management.}
KVFlow~\citep{pang2025kvflow} predicts future cache needs from workflow-graph structure; Marconi~\citep{pan2025marconi} forecasts reuse from prior traffic. PEEK shares the goal but derives demand from the live pending queue via direct reference counting---no execution graph, no learned forecast---and is complementary across within-tick (PEEK) and across-tick (KVFlow/Marconi) horizons.

%% file: content/conclusion.tex
\section{Conclusion}
\vspace{-0.4em}
\textbf{PEEK} treats the waiting queue as a first-class signal for KV cache management via an incremental radix tree, a dual-walk yielding all-LPM in $O(C{\times}D)$, cluster-aware admission/eviction, and an adaptive fairness lane. Across five workloads on SGLang/vLLM up to 4$\times$H100, PEEK lifts cache hit $\mathbf{3.0\times/2.6\times}$, TTFT $\mathbf{7.9\times/7.1\times}$, E2E $\mathbf{6.7\times/5.5\times}$, throughput $\mathbf{3.6\times/4.5\times}$, matches baselines on no-sharing traffic, and composes beneath cross-replica routers and SLO arbiters atop any prefix-reuse mechanism. \textit{Limitations:} the $N/C$ amortization shrinks toward 1 in singleton-dominated traffic (the \texttt{has\_sharing} guard handles this regime); the cLPM sort still runs in Python; evaluation is on H100 only, leaving smaller-memory accelerators and multi-modal prefixes for future work. Code will be released.

%% file: content/appendix.tex
\section{PEEK: Predictive Queue-Informed KV Cache Management Mechanisms}
\label{app:peek_mechanisms}

\subsection{PEEK vs.\ SGLang LPM: Mechanism Comparison}
\label{app:peek_vs_lpm}

PEEK keeps stock LPM's longest-prefix-match objective and reuses its pioneer/sibling threshold ($\theta{=}32$ tokens), but otherwise replaces every other component with a pending-tree-driven counterpart. Table~\ref{tab:peek-vs-lpm} lays out the differences axis by axis; each row maps to a section of the main text or this appendix.

\begin{table}[!htb]
\centering
\small
\setlength{\tabcolsep}{4pt}
\renewcommand{\arraystretch}{1.15}
\begin{tabular}{p{0.18\textwidth} p{0.34\textwidth} p{0.40\textwidth}}
\toprule
\textbf{Axis} & \textbf{SGLang LPM (stock)} & \textbf{PEEK (cLPM + PE + DL)} \\
\midrule
Prefix signal source &
Engine-side radix cache only --- \emph{reactive}: ranks by what is currently cached &
Engine cache \emph{plus} a pending-queue radix tree --- \emph{predictive}: ranks by what waiting requests will share (\S\ref{sec:pending_tree}) \\
\addlinespace[2pt]
Scheduler sort key &
$(-h(r),\, \texttt{is\_depr}(r),\, \texttt{arrival}(r))$ &
$(\texttt{section}(r),$ $-h(r),$ $-\texttt{score}(r),$ $-|\texttt{cluster}(r)|,$ $\texttt{arrival}(r))$ (\S\ref{sec:clpm_sched}) \\
\addlinespace[2pt]
Cold-tie behavior ($h{=}0$) &
Falls back to arrival (FCFS) --- interleaves clusters, induces thrashing (\S\ref{sec:thrashing}) &
Largest-cluster pioneer first; siblings sort directly behind, so one prefill amortizes across the whole cluster \\
\addlinespace[2pt]
In-batch deprioritization &
Auxiliary radix tree rebuilt from scratch each scheduling call (throwaway) &
Read off the persistent pending tree at $O(1)$ per request --- no rebuild \\
\addlinespace[2pt]
Per-cycle LPM cost &
$O(N{\times}D)$ per-request cache lookups &
$O(C{\times}D)$ via tree-vs-tree dual-walk on SGLang ($C$ = \#distinct shared prefixes, $C{\ll}N$ in high-sharing queues); reuses engine probe on vLLM (\S\ref{sec:dual_walk}) \\
\addlinespace[2pt]
Cluster signals exposed &
None &
Cluster identity, cluster size, depth-weighted ancestor demand score \\
\addlinespace[2pt]
Eviction policy &
LRU/LFU on cache state only --- decoupled from scheduling &
Queue-aware hook scoring blocks by $\max_{v \in \text{ancestors}}\!\texttt{pending\_count}(v){\cdot}\texttt{depth}(v)$: zero-demand blocks first, blocks the largest pending clusters depend on protected last (\S\ref{sec:eviction}) \\
\addlinespace[2pt]
Starvation handling &
None beyond the arrival tiebreak &
Stride-interleaved fairness lane $(\texttt{section},\, \texttt{arrival},\, -h)$; a dynamic-lane controller widens the fairness share with singleton fraction and oldest-singleton wait, EMA-smoothed (\S\ref{sec:clpm_sched}) \\
\addlinespace[2pt]
No-sharing fallback &
Always pays the LPM/aux-tree cost &
\texttt{has\_sharing} guard short-circuits to FCFS at zero overhead when the queue contains no shared prefixes (\S\ref{sec:guards}) \\
\addlinespace[2pt]
Scheduling/eviction coupling &
Decoupled; eviction cannot anticipate scheduler choices &
Both consume one shared signal (the pending tree) --- co-designed without explicit coordination \\
\bottomrule
\end{tabular}
\caption[PEEK vs.\ SGLang LPM: mechanism-level comparison]{PEEK vs.\ stock SGLang LPM, axis by axis. PEEK preserves LPM's longest-prefix-match objective and pioneer/sibling threshold but replaces FCFS tiebreaks with cluster-aware ordering, eliminates the per-cycle aux-tree rebuild via the persistent pending tree, drives queue-aware eviction off the same tree, and bounds singleton starvation with an adaptive fairness lane.}
\label{tab:peek-vs-lpm}
\end{table}

\subsection{Implementation Details}
\label{app:implementation}

\paragraph{Native core.} The pending radix tree, dual-walk, and per-rid cluster/score lookups are implemented in $\sim$1.6\,kLOC of Rust (\texttt{src/tree.rs}, \texttt{src/pending.rs}) and exposed to Python via PyO3 (\texttt{peek.\_core}). Nodes live in a contiguous arena indexed by \texttt{NodeId}, with \texttt{FxHashMap} children keyed by the first edge token, \texttt{FxHashSet} terminators, and per-node \texttt{pending\_count}; \texttt{insert}/\texttt{remove} run in $O(D)$ with no allocation in the hot path. The cLPM sort itself runs in Python ($O(N \log N)$ over the Rust-computed signals); pushing the tree, dual-walk, and per-rid score/cluster computations into Rust eliminates the per-compare Python callback overhead that dominates stock SGLang LPM at large $N$.

\paragraph{SGLang integration ($\sim$814 LOC of Python).} \texttt{peek.engines.sglang.patch\_hook} monkey-patches two surfaces at import time. \texttt{RadixCache.\_\_init\_\_} is patched to install \texttt{PeekDemandStrategy} (\S\ref{sec:eviction}) in place of the default heap-based eviction. \texttt{Scheduler.process\_input\_requests} is patched to diff-sync the pending tree against \texttt{self.waiting\_queue} on every scheduling iteration---a mutation-path-agnostic sync that tolerates any downstream queue manipulation (append, pop, list-comprehension reassignment) without further hooks. The same hook overrides \texttt{SchedulePolicy.calc\_priority} (the entry point that internally calls \texttt{\_sort\_by\_longest\_prefix} in stock SGLang) to invoke PEEK's cLPM sort.

\paragraph{vLLM integration ($\sim$800 LOC of Python).} \texttt{peek.engines.vllm.patch\_hook} hooks three v1 scheduler entry points: \texttt{Scheduler.schedule} (sync + reorder \texttt{self.waiting}), \texttt{Scheduler.add\_request} (sync-on-arrival fast path), and \texttt{Scheduler.finish\_requests} (drop completed rids). Because vLLM's prefix cache is hash-keyed rather than tree-structured, PEEK additionally maintains an inverted index $\text{block\_hash} \to \text{demand}$ alongside the pending tree, and patches \texttt{BlockPool.get\_new\_blocks} to scan the free-block queue (capped at $4N$ entries) and prefer zero-demand victims over the default LRU pick. Recency/decay eviction modes also patch \texttt{BlockPool.touch} and \texttt{cache\_full\_blocks} to fabricate per-block timestamps that vLLM does not track natively. vLLM v1 spawns its \texttt{EngineCore} via \texttt{multiprocessing.spawn}, so monkey-patches in the parent do not inherit; PEEK ships a \texttt{sitecustomize.py} shim that re-imports the hook in every child interpreter.

\paragraph{Activation surface.} A single environment flag selects scope: \texttt{PEEK\_ONLINE\_SCHEDULER} for cluster-aware scheduling, \texttt{PEEK\_ONLINE\_EVICTION} for queue-aware eviction, \texttt{PEEK\_ONLINE\_CLPM} for the multi-lane scheduler, \texttt{PEEK\_ONLINE\_EVICTION\_MODE} $\in$ \{\texttt{plain}, \texttt{cluster}, \texttt{recency}, \texttt{decay}\} for the eviction variant, and \texttt{PEEK\_ONLINE\_CLPM\_BIGLANE\_SHARE} for the cache-locality lane share (default 0.7). With no flags set, both shims are no-ops and the engine runs unmodified. Because integration is by monkey-patching, no fork or PR against either upstream engine is required, and PEEK tracks SGLang's main branch and vLLM 0.19.1 with byte-identical baselines.

\subsection{Pending Radix Tree: Build and Clustering}
\label{app:pending_tree_alg}

The pending radix tree (\S\ref{sec:pending_tree}) is a textbook compressed trie with two augmentations: per-node \texttt{pending\_count} and per-leaf \texttt{terminators}. \textsc{Insert} descends from the root following matching edges, splitting any edge on token divergence (creating a new internal node) and adding a new leaf for the request, then bumps \texttt{pending\_count} along the root-to-leaf path---$O(D)$, where $D$ is the prompt's tree depth. \textsc{Remove} reverses: walk to the request's terminator, remove the rid, decrement \texttt{pending\_count} along the path, and prune empty branches---also $O(D)$. \textsc{Cluster} walks the request's prompt down the tree and returns the deepest node $v$ with $\texttt{pending\_count}(v) \geq 2$; cluster size is $\texttt{pending\_count}(v)$. The tree therefore costs $O(N \cdot D)$ per cycle in total, dominated in practice by the scheduler's $O(N \log N)$ sort (\S\ref{sec:clpm_sched}).

\subsection{Dual-Walk for Longest Prefix Match}
\label{app:dual_walk_alg}

Figure~\ref{fig:dualwalk} contrasts the two engines. On SGLang's radix cache, PEEK runs the recursive dual-walk: tree-vs-tree co-descent collapses $N$ per-request lookups into $C$ per-cluster lookups (Algorithm~\ref{alg:dual_walk}). The recursion descends a pending edge only when the edge matches in full---a partial match terminates the branch, since deeper nodes cannot extend the cached prefix. On vLLM's flat block-hash cache, no structural co-descent is possible: PEEK calls the engine's per-request \texttt{find\_longest\_cache\_hit} once per waiting request to obtain $h(r)$, then walks the pending tree separately to read cluster size and ancestor score---the LPM lookup is not amortized, but the cluster signal stock vLLM's FCFS lacks is still surfaced.

\begin{figure}[t]
\centering
\begin{subfigure}[b]{\columnwidth}
\centering
\resizebox{0.85\linewidth}{!}{%
\begin{tikzpicture}[
  pnode/.style={circle, draw=orange!60!black, fill=orange!10, inner sep=1pt, font=\scriptsize, minimum size=4mm},
  cnode/.style={circle, draw=red!50!black, fill=red!8, inner sep=1pt, font=\scriptsize, minimum size=4mm},
  leaf/.style={font=\tiny, text=gray!40!black},
  edge/.style={->, >={Stealth[length=1.4mm]}, thin},
  match/.style={->, >={Stealth[length=1.6mm]}, dashed, thick, gray!60!black},
]
  \node[font=\scriptsize\bfseries] at (0, 2.4) {Pending Tree};
  \node[pnode] (proot) at (0, 1.8) {};
  \node[pnode] (pA) at (-1.0, 0.9) {3};
  \node[pnode] (pB) at (1.0, 0.9) {2};
  \node[leaf, below=0pt of pA, anchor=north] {A1,A2,A3};
  \node[leaf, below=0pt of pB, anchor=north] {B1,B2};
  \draw[edge] (proot) -- node[font=\tiny, midway, above left=-2pt] {sys+A} (pA);
  \draw[edge] (proot) -- node[font=\tiny, midway, above right=-2pt] {sys+B} (pB);
  \node[font=\scriptsize\bfseries] at (5, 2.4) {SGLang Radix Cache};
  \node[cnode] (croot) at (5, 1.8) {};
  \node[cnode] (csys) at (5, 0.9) {};
  \node[leaf, right=2pt of csys] {sys (cached)};
  \draw[edge, red!50!black] (croot) -- node[font=\tiny, midway, right=-1pt] {sys} (csys);
  \draw[match] (proot.east) -- node[font=\tiny\itshape, midway, above] {co-descend} (croot.west);
  \draw[match] (pA.east) -- node[font=\tiny, midway, above, sloped] {1 query $\to$ 3 hits} ([yshift=-2pt]csys.west);
  \draw[match] (pB.east) -- node[font=\tiny, midway, below, sloped] {1 query $\to$ 2 hits} ([yshift=-6pt]csys.west);
  \node[font=\tiny, anchor=west] at (-1.6, -0.5) {Stock LPM: 5 queries ($O(N{\times}D)$). \quad PEEK: 2 queries ($O(C{\times}D)$).};
\end{tikzpicture}}
\caption{Dual-walk on SGLang: tree-vs-tree co-descent amortizes lookups across cluster siblings.}
\label{fig:dualwalk_sglang}
\end{subfigure}

\vspace{0.6em}

\begin{subfigure}[b]{\columnwidth}
\centering
\resizebox{0.85\linewidth}{!}{%
\begin{tikzpicture}[
  pnode/.style={circle, draw=orange!60!black, fill=orange!10, inner sep=1pt, font=\scriptsize, minimum size=4mm},
  hcell/.style={rectangle, draw=red!50!black, fill=red!8, font=\tiny, minimum width=15mm, minimum height=3.5mm, inner sep=1pt},
  leaf/.style={font=\tiny, text=gray!40!black},
  edge/.style={->, >={Stealth[length=1.4mm]}, thin},
  probe/.style={->, >={Stealth[length=1.4mm]}, dashed, thin, gray!50!black},
  cluster/.style={->, >={Stealth[length=1.6mm]}, thick, orange!70!black},
]
  \node[font=\scriptsize\bfseries] at (0, 2.4) {Pending Tree};
  \node[pnode] (proot) at (0, 1.8) {};
  \node[pnode] (pA) at (-1.0, 0.9) {3};
  \node[pnode] (pB) at (1.0, 0.9) {2};
  \node[leaf, below=0pt of pA, anchor=north] {A1,A2,A3};
  \node[leaf, below=0pt of pB, anchor=north] {B1,B2};
  \draw[edge] (proot) -- node[font=\tiny, midway, above left=-2pt] {sys+A} (pA);
  \draw[edge] (proot) -- node[font=\tiny, midway, above right=-2pt] {sys+B} (pB);
  \node[font=\scriptsize\bfseries] at (5, 2.4) {vLLM Block-Hash Cache};
  \node[hcell] (h1) at (5, 1.85) {h(sys blk\textsubscript{0})};
  \node[hcell] (h2) at (5, 1.45) {h(sys blk\textsubscript{1})};
  \node[hcell] (h3) at (5, 1.05) {h(sys+A blk)};
  \node[hcell] (h4) at (5, 0.65) {\ldots};
  \draw[probe] (pA.east) -- node[font=\tiny, midway, above, sloped] {3 probes} (h3.west);
  \draw[probe] (pB.east) -- node[font=\tiny, midway, below, sloped] {2 probes} ([yshift=-2pt]h3.west);
  \draw[cluster] (proot.south) ++(0, -0.05) -- ++(0, -0.6) -| ++(2.0, -0.4) node[font=\tiny, midway, above] {tree walk $\to$ cluster signal};
  \node[font=\tiny, anchor=west] at (-1.6, -0.5) {Stock vLLM: 5 hash probes, no cluster signal. \quad PEEK: 5 probes + tree walk surfaces clusters for cLPM.};
\end{tikzpicture}}
\caption{Dual-walk on vLLM: per-request hash probes are unavoidable, but the pending-tree walk surfaces the cluster signal that vLLM's flat hash cache cannot.}
\label{fig:dualwalk_vllm}
\end{subfigure}
\caption{Dual-walk on (a) SGLang and (b) vLLM. SGLang's radix cache permits tree-vs-tree co-descent, collapsing $N$ per-request LPM lookups into $C$ per-cluster lookups. vLLM's flat block-hash cache forces one probe per request, but the pending-tree walk still extracts cluster information for the scheduler downstream.}
\label{fig:dualwalk}
\end{figure}

\textbf{Cost.} On SGLang, each pending edge incurs one cache co-traversal step; total is $O(C \times D)$ across $C$ pending tree edges of average depth $D$, vs.\ $O(N \times D)$ for stock LPM---an amortization factor of $N/C$ (large when sharing is high). On vLLM, the LPM lookup itself is not amortized: $O(N \times b)$ for $N$ \texttt{find\_longest\_cache\_hit} calls ($b$ = block-hash chain length), matching stock APC's per-request cost; the additional $O(N)$ pending-tree walk for cluster signals is dwarfed by the LPM probes. Algorithm~\ref{alg:dual_walk} (\S\ref{sec:dual_walk}) details the SGLang procedure.

\subsection{cLPM Scheduling Algorithm}
\label{app:clpm_algorithm}

Algorithm~\ref{alg:clpm} details the per-cycle scheduling routine described in \S\ref{sec:clpm_sched}: dual-walk (or per-request LPM on vLLM) for $h(r)$, pending-tree lookups for cluster signals, lexicographic sort of two independent lane keys (cache-locality lane keyed by Eq.~\ref{eq:clpm_key_a}; fairness lane keyed by arrival), and stride-interleave between them.

\begin{algorithm}[t]
\caption{cLPM Per-Cycle Scheduling}
\label{alg:clpm}
\begin{algorithmic}[1]
\REQUIRE Waiting queue $\mathcal{R}$, pending tree $T$, engine cache $\mathcal{C}$, cache-lane share $\alpha$, warm threshold $\tau$, deprio threshold $\theta$
\ENSURE Admission order over $\mathcal{R}$
\STATE $H \leftarrow \texttt{dual\_walk}(T, \mathcal{C})$ \COMMENT{$H[r] = h(r)$ for each $r$ (Alg.~\ref{alg:dual_walk})}
\STATE $S \leftarrow \texttt{compute\_req\_scores}(T)$;\quad $K \leftarrow \texttt{all\_cluster\_info}(T)$ \COMMENT{batched Rust calls}
\STATE $\texttt{seen} \leftarrow \emptyset$
\FOR{each $r \in \mathcal{R}$ in arrival order}
    \IF{$H[r] > \tau$} \STATE $\texttt{section}(r) \leftarrow 0$ \COMMENT{warm}
    \ELSE
        \STATE $p \leftarrow \texttt{prompt}(r)[{:}\theta]$
        \IF{$p \in \texttt{seen}$} \STATE $\texttt{section}(r) \leftarrow 2$ \COMMENT{cold sibling (deprio tail)}
        \ELSE \STATE $\texttt{section}(r) \leftarrow 1$;\;\; $\texttt{seen} \leftarrow \texttt{seen} \cup \{p\}$ \COMMENT{cold pioneer} \ENDIF
    \ENDIF
    \STATE $\texttt{score}(r) \leftarrow S[r]$;\quad $|\texttt{cluster}(r)| \leftarrow K[r].\texttt{size}$
\ENDFOR
\STATE $\mathcal{R}_A \leftarrow \texttt{sort}(\mathcal{R}, \text{key}=\text{Eq.~\ref{eq:clpm_key_a}})$ \COMMENT{cache-locality lane}
\STATE $\mathcal{R}_B \leftarrow \texttt{sort}(\mathcal{R}, \text{key}=(\texttt{section}, \texttt{arrival}, -H))$ \COMMENT{fairness lane}
\IF{\texttt{dynamic\_lane}}
    \STATE $\alpha \leftarrow 1 - \texttt{EMA}_{\beta=0.3}\!\big(\texttt{clamp}(0.15 + 0.5\cdot\texttt{singleton\_frac} + 0.3\cdot\texttt{age\_pressure},\, 0.1,\, 0.6)\big)$
\ENDIF
\STATE \textbf{return} $\texttt{stride\_interleave}(\mathcal{R}_A, \mathcal{R}_B, \alpha)$ \COMMENT{cache-locality lane gets share $\alpha$ (default 0.7)}
\end{algorithmic}
\end{algorithm}

\textbf{Cost.} Dual-walk: $O(C \times D)$ on SGLang. On vLLM, the cluster signal is read directly from the pending tree via \texttt{all\_cluster\_info} ($O(N)$) and per-request LPM is satisfied by one $O(b)$ probe each into the engine's block-hash index ($b$ = block-hash chain length), totaling $O(N \times b)$ for the LPM lookup; the structural pending-tree walk used by SGLang is unnecessary because vLLM exposes a direct flat-cache lookup. Per-request \texttt{score}/\texttt{cluster} reads are $O(1)$ (computed once via batched Rust calls). Sort: $O(N \log N)$. Stride interleave: $O(N)$. Total per cycle: $O(C \times D + N \log N)$ on SGLang; $O(N \times b + N \log N)$ on vLLM.

\textbf{Guard.} The pending tree exposes \texttt{has\_sharing()} (an $O(\#\text{root-children})$ check); when integration is configured to skip cLPM on no-sharing queues, Lines 1--16 are bypassed and the engine's stock order is returned. Empirically PEEK matches FCFS within $\pm$1.5\% on no-sharing workloads (\S\ref{sec:guards}, \S\ref{sec:noregress}), driven by both this guard and SGLang's stock LPM-fallback at queue depth $>$128.

\subsection{Eviction Motivation Example}
\label{app:eviction}

Even with good scheduling, uninformed LRU eviction can undo prefix sharing. Consider three prefix groups A, B, C with a cache holding two prefixes. Wave~1 processes A then B, leaving both cached (B most recently used). Wave~2 needs A and C. When C arrives and the cache is full, LRU evicts A---the least recently accessed---but A is exactly what wave~2 needs next, forcing full recomputation. A queue-aware policy inspects pending requests, sees A is needed and B is not, and evicts B instead---preserving A for an immediate hit.

\begin{figure}[ht]
\centering
\begin{subfigure}[b]{0.48\textwidth}
    \centering
    \begin{tikzpicture}[scale=0.55, every node/.style={font=\scriptsize}]
        \node[font=\scriptsize\bfseries, anchor=west] at (0, 5.2) {Wave 1};
        \node[font=\scriptsize\bfseries, anchor=west] at (5.5, 5.2) {Wave 2};
        \fill[red!60] (0, 4.4) rectangle ++(1.0, 0.6); \node at (0.5, 4.7) {A};
        \fill[blue!60] (1.2, 4.4) rectangle ++(1.0, 0.6); \node at (1.7, 4.7) {B};
        \fill[red!60] (5.5, 4.4) rectangle ++(1.0, 0.6); \node at (6.0, 4.7) {A};
        \fill[green!60] (6.7, 4.4) rectangle ++(1.0, 0.6); \node at (7.2, 4.7) {C};
        \node[font=\scriptsize\bfseries, anchor=east] at (-0.2, 3.2) {Cache:};
        \node[font=\tiny, anchor=south] at (1.5, 3.7) {after wave 1};
        \fill[red!40] (0.5, 2.9) rectangle ++(0.8, 0.5); \node[font=\tiny] at (0.9, 3.15) {A};
        \fill[blue!40] (1.5, 2.9) rectangle ++(0.8, 0.5); \node[font=\tiny] at (1.9, 3.15) {B};
        \draw[-{Stealth[length=1.5mm]}, thick, red!70] (3.0, 3.15) -- node[above, font=\tiny] {evict A} (4.2, 3.15);
        \node[font=\tiny, anchor=south] at (5.5, 3.7) {after eviction};
        \fill[blue!40] (4.5, 2.9) rectangle ++(0.8, 0.5); \node[font=\tiny] at (4.9, 3.15) {B};
        \fill[green!40] (5.5, 2.9) rectangle ++(0.8, 0.5); \node[font=\tiny] at (5.9, 3.15) {C};
        \node[font=\tiny, red!70] at (7.2, 3.15) {A: \textbf{miss}};
        \node[font=\scriptsize, anchor=west] at (0, 2.1) {A must be recomputed from scratch};
    \end{tikzpicture}
    \caption{LRU evicts A (least recent), but A is needed next.}
    \label{fig:eviction_lru}
\end{subfigure}
\hfill
\begin{subfigure}[b]{0.48\textwidth}
    \centering
    \begin{tikzpicture}[scale=0.55, every node/.style={font=\scriptsize}]
        \node[font=\scriptsize\bfseries, anchor=west] at (0, 5.2) {Wave 1};
        \node[font=\scriptsize\bfseries, anchor=west] at (5.5, 5.2) {Wave 2};
        \fill[red!60] (0, 4.4) rectangle ++(1.0, 0.6); \node at (0.5, 4.7) {A};
        \fill[blue!60] (1.2, 4.4) rectangle ++(1.0, 0.6); \node at (1.7, 4.7) {B};
        \fill[red!60] (5.5, 4.4) rectangle ++(1.0, 0.6); \node at (6.0, 4.7) {A};
        \fill[green!60] (6.7, 4.4) rectangle ++(1.0, 0.6); \node at (7.2, 4.7) {C};
        \node[font=\scriptsize\bfseries, anchor=east] at (-0.2, 3.2) {Cache:};
        \node[font=\tiny, anchor=south] at (1.5, 3.7) {after wave 1};
        \fill[red!40] (0.5, 2.9) rectangle ++(0.8, 0.5); \node[font=\tiny] at (0.9, 3.15) {A};
        \fill[blue!40] (1.5, 2.9) rectangle ++(0.8, 0.5); \node[font=\tiny] at (1.9, 3.15) {B};
        \draw[-{Stealth[length=1.5mm]}, thick, green!50!black] (3.0, 3.15) -- node[above, font=\tiny] {evict B} (4.2, 3.15);
        \node[font=\tiny, anchor=south] at (5.5, 3.7) {after eviction};
        \fill[red!40] (4.5, 2.9) rectangle ++(0.8, 0.5); \node[font=\tiny] at (4.9, 3.15) {A};
        \fill[green!40] (5.5, 2.9) rectangle ++(0.8, 0.5); \node[font=\tiny] at (5.9, 3.15) {C};
        \node[font=\tiny, green!50!black] at (7.2, 3.15) {A: \textbf{hit}};
        \node[font=\scriptsize, anchor=west] at (0, 2.1) {A is preserved for immediate reuse};
    \end{tikzpicture}
    \caption{Queue-aware: evicts B (not needed), preserves A.}
    \label{fig:eviction_aware}
\end{subfigure}
\caption{LRU vs.\ queue-aware eviction. After wave~1 caches A and B, wave~2 needs A and C. LRU (a) evicts A---the least recently used---forcing recomputation. Queue-aware eviction (b) inspects pending requests, preserves A, and evicts the unreferenced B.}
\label{fig:eviction}
\end{figure}

\section{PEEK Experiments}
\label{app:peek_experiments}

\subsection{Experimental Setup Tables}
\label{app:exp_tables}

\begin{table}[ht]
\caption{Hardware, model, and dataset assignment per workload.}
\label{tab:hardware}
\centering
\footnotesize
\setlength{\tabcolsep}{4pt}
\begin{tabular}{p{0.04\linewidth} p{0.22\linewidth} p{0.27\linewidth} p{0.40\linewidth}}
\toprule
\textbf{W} & \textbf{GPUs} & \textbf{Model (bf16)} & \textbf{Dataset} \\
\midrule
W1 & 1$\times$NVIDIA H100 80\,GB        & Qwen2.5-32B-Instruct                                       & LooGLE long-document corpus: 100 group prompts ($\sim$1024 tokens each), Zipf-$\alpha$=1.0 sampling \\
\addlinespace
W2 & 1$\times$NVIDIA H100 80\,GB        & Qwen2.5-32B-Instruct                                       & RepoBench-Python v1.1 (\texttt{cross\_file\_first} split): 40 repository contexts truncated to $\sim$8192-token shared prefixes, Zipf-$\alpha$=1.0 sampling \\
\addlinespace
W3 & 2--4$\times$NVIDIA H100 80\,GB (TP=2; DP=1 / DP=2) & Llama-3.1-70B-Instruct                  & Multi-GPU scaling of W1 and W2: cell B mirrors W2 (long-prefix RAG), cell C mirrors W1 (shared-prompt chat) \\
\addlinespace
W4 & 1$\times$NVIDIA H100 80\,GB        & Mistral-Small-24B-Instruct-2501 (matches Mooncake trace)   & Mooncake \texttt{conversation\_trace} (FAST'25) with within-session prefix accumulation; \emph{agentic\_only} and \emph{agentic\_shared} scenarios \\
\addlinespace
W5 & 1$\times$NVIDIA H100 80\,GB        & Gemma-2-27B-it                                             & LMSYS arena chat (heterogeneous singletons): \emph{cell\_C\_long} (512--4096 tokens) and \emph{cell\_C\_short} (32--1024 tokens) \\
\bottomrule
\end{tabular}
\end{table}

\begin{table}[ht]
\caption{Shared experiment setup across W1--W5.}
\label{tab:setup}
\centering
\footnotesize
\setlength{\tabcolsep}{4pt}
\begin{tabular}{p{0.20\linewidth} p{0.74\linewidth}}
\toprule
\textbf{Component} & \textbf{Configuration} \\
\midrule
Engines           & SGLang 0.5.9, vLLM 0.19.1 \\
Framework         & PyTorch 2.9.1, Python 3.12.13, CUDA graphs \emph{on} \\
Memory budget     & SGLang \texttt{mem\_fraction\_static}=0.88; vLLM \texttt{gpu\_memory\_utilization}=0.9 \\
Seeds             & 42, 142, 242 (3 replicates per cell; reported numbers are means across seeds; $\pm$ values are seed standard deviation) \\
Warmup            & First $\sim$10--20\% of requests per cell (50--400 requests depending on cell size) are excluded from metric aggregation \\
Load levels        & Two operating points per cell, calibrated by sustained queue depth: \emph{moderate} maintains a queue length of $\sim$60--100 requests; \emph{heavy} sustains $\sim$150--200 \\
Policy lattice (SGLang) & B0 (FCFS+LRU), B1 (LPM+LRU); \texttt{P\_clpm} (cLPM only), \texttt{P\_full} (cLPM + group-major admission + queue-aware eviction) \\
Policy lattice (vLLM)   & B2 (FCFS+APC+LRU, strongest stock baseline); P2 (eviction only), P3 (cLPM only), P4 (cLPM + group-major), P5 (P4 + dynamic-lane fairness), P6 (P4 + queue-aware eviction, \textbf{primary co-design}), P7 (P5 + queue-aware eviction) \\
\bottomrule
\end{tabular}
\end{table}

\begin{table}[ht]
\caption{Experiments and per-workload configurations. ``Probe axis'' names the design facet each workload stresses relative to the W1 driving scenario.}
\label{tab:per_workload}
\centering
\footnotesize
\setlength{\tabcolsep}{3pt}
\begin{tabular}{p{0.04\linewidth} p{0.16\linewidth} p{0.20\linewidth} p{0.50\linewidth}}
\toprule
\textbf{W} & \textbf{Probe axis} & \textbf{Cells} & \textbf{Configuration} \\
\midrule
W1 & Oversubscription (driving) & A: 2$\times$, B: 4$\times$, C: 8$\times$ (primary), D: 16$\times$ KV pressure; moderate / heavy & Zipf-$\alpha$=1.0 prefix groups, Poisson arrivals, fixed decode=128. Full ablation lattice (B2, P2--P7). \\
\addlinespace
W2 & Long, variable decode (decode-dominant) & cell B (4$\times$), moderate / heavy & 40 documents $\times$ 8192-token prefixes; decode mix \texttt{10:128, 25:512, 30:1024, 25:2048, 10:4096} (mean $\approx$1460); KV footprint $\approx$7$\times$ arena, eviction always firing. \\
\addlinespace
W3 & Multi-GPU scaling (Llama-3.1-70B, TP=2) & cell B (mirrors W2, RAG-like) and cell C (mirrors W1, chat-like) at DP=1 (2$\times$H100) and DP=2 (4$\times$H100); heavy & Same dataset patterns as W1 (cell C) and W2 (cell B), executed at 70B scale across two data-parallel topologies. \\
\addlinespace
W4 & Within-session prefix accumulation (Mooncake) & \texttt{agentic\_only}, \texttt{agentic\_shared}; moderate / heavy & Mooncake \texttt{conversation\_trace} (FAST'25), 4 rounds, 50\,ms inter-turn. \texttt{agentic\_shared} adds a 1402-token shared system prompt (Cursor / Copilot / Claude Code pattern). \\
\addlinespace
W5 & Singleton-dominated chat (no-regress) & \texttt{cell\_C\_long} (512--4096 tokens, weak residual overlap), \texttt{cell\_C\_short} (32--1024 tokens, no structural sharing); heavy & LMSYS arena chat singletons; lognormal length distribution; decode 64--256; Poisson arrivals; $N$=1500. \\
\bottomrule
\end{tabular}
\end{table}

\begin{table}[ht]
\caption{W1 full configuration. Top: shared across all cells. Bottom: per-cell parameters and per-engine calibrated arrival rates. KV footprint is the peak working-set size; oversubscription is footprint relative to the $\sim$50K-token arena. Rates target moderate $=$ queue p50 in [60,127], heavy $\geq$ 128 (the SGLang LPM-fallback boundary).}
\label{tab:w1_cells}
\centering
\footnotesize
\setlength{\tabcolsep}{4pt}

\begin{tabular}{p{0.20\linewidth} p{0.74\linewidth}}
\toprule
\multicolumn{2}{c}{\textbf{Shared across all W1 cells}} \\
\midrule
Hardware           & 1$\times$NVIDIA H100 80\,GB (bf16) \\
Model              & Qwen2.5-32B-Instruct \\
Engines            & SGLang 0.5.9 (B1=LPM+LRU baseline); vLLM 0.19.1 (B2=FCFS+APC+LRU baseline) \\
Dataset            & LooGLE long-document corpus, prefix tokens taken from documents \\
Group sampling     & Zipf-$\alpha$=1.0 over $G$ shared-prompt groups \\
Arrival process    & Poisson (inter-arrival $\sim$Exp($\lambda$)) \\
Decode             & Fixed 128 tokens per request \\
Concurrency cap    & vLLM: moderate=100, heavy=210; SGLang: 256 \\
Seeds              & 42, 142, 242 (3 replicates per cell; reported numbers are means across seeds) \\
Warmup             & First $N \cdot 0.2$ requests excluded from metrics \\
Policies           & B1 (SGLang) / B2 (vLLM); P2 (eviction-only); P3 (cLPM); P4 (cLPM+group-major); P5 (P4+dyn-lane); \textbf{P6} (P4+queue-aware eviction, primary co-design); P7 (P5+queue-aware eviction) \\
\bottomrule
\end{tabular}

\vspace{0.5em}
\begin{tabular}{lccccccccccc}
\toprule
& & \textbf{prefix} & & \textbf{warmup} & \textbf{KV} & \textbf{over-} & \multicolumn{2}{c}{\textbf{SGLang (req/s)}} & \multicolumn{2}{c}{\textbf{vLLM (req/s)}} \\
\cmidrule(lr){8-9}\cmidrule(lr){10-11}
\textbf{Cell} & \textbf{$G$} & \textbf{(tok)} & \textbf{$N$} & \textbf{(req)} & \textbf{(tok)} & \textbf{sub.} & \textbf{mod.} & \textbf{heavy} & \textbf{mod.} & \textbf{heavy} \\
\midrule
A          & 100 & 1024 & 1000 & 200 & 102\,K & 2$\times$  & 40 & 60 & 20 & 30 \\
B          & 200 & 1024 & 2000 & 400 & 205\,K & 4$\times$  & 35 & 80 & 20 & 30 \\
\textbf{C} & \textbf{100} & \textbf{4096} & \textbf{1000} & \textbf{200} & \textbf{410\,K} & \textbf{8$\times$ (primary)} & \textbf{4} & \textbf{8} & \textbf{5} & \textbf{10} \\
D          & 200 & 4096 & 2000 & 400 & 820\,K & 16$\times$ & 12 & 24 & 8  & 15 \\
\bottomrule
\end{tabular}
\end{table}

\begin{table}[ht]
\caption{W1 cells mapped to production deployment classes. Cell C is designated primary as the most directly representative of PEEK's target deployment.}
\label{tab:w1_cell_meaning}
\centering
\footnotesize
\setlength{\tabcolsep}{4pt}
\begin{tabular}{p{0.04\linewidth} p{0.32\linewidth} p{0.58\linewidth}}
\toprule
\textbf{Cell} & \textbf{Production analog} & \textbf{Role / question this cell answers} \\
\midrule
A (2$\times$)  & SMB chatbot, single-tenant SaaS (e.g.\ Intercom Fin AI) where the KV cache mostly fits & ``Does PEEK matter at all when KV mostly fits?'' --- near-baseline reference \\
\addlinespace
B (4$\times$)  & Multi-tenant SaaS / mid-size prompt library (Jasper-style content templates; multi-vertical chat platforms aggregating $\sim$200 small tenants) & ``Does PEEK help the typical SaaS deployment in steady state?'' --- common-case production test \\
\addlinespace
\textbf{C (8$\times$, primary)} & \textbf{Complex prompts at scale}: ChatGPT custom GPTs, Claude projects with tool-using assistants, Cursor / Copilot workspace prompts with file/repo context, enterprise platforms with $\sim$100 tool-using personas & \textbf{``Does PEEK help ChatGPT-class deployments under realistic peak pressure?''} --- the workload the design directly targets \\
\addlinespace
D (16$\times$) & Large multi-tenant agent platform under flash / extreme load: launch-day API peak, viral-spike chat platform, multi-vertical agent platform during launch & ``Does PEEK hold up under flash / overload, or collapse along with everything else?'' --- stress test \\
\bottomrule
\end{tabular}
\end{table}

\subsection{Workload-to-Production-Scenario Mapping}
\label{app:prod_mapping}

\begin{table}[ht]
\caption{W1--W5 mapping to production deployment classes and serving hardware.}
\label{tab:prod_mapping}
\centering
\footnotesize
\setlength{\tabcolsep}{4pt}
\begin{tabular}{p{0.05\linewidth} p{0.22\linewidth} p{0.50\linewidth} p{0.13\linewidth}}
\toprule
\textbf{W} & \textbf{Class} & \textbf{Production scenarios} & \textbf{GPUs} \\
\midrule
W1 & Shared-prompt / templated apps     & SMB chatbots (Intercom Fin), multi-tenant SaaS prompt libraries (Jasper-style), custom-GPT / Claude-projects at scale, enterprise multi-tenant agent platforms under flash load & 1$\times$H100 \\
\addlinespace
W2 & Long-document RAG (decode-dominant) & Enterprise RAG over document corpora (legal, medical, finance); long-form chat-over-doc; reasoning-adjacent synthesis & 1$\times$H100 \\
\addlinespace
W3 & Multi-GPU 70B serving for shared-prompt + RAG & Production deployments of W1- and W2-class workloads on Llama-3.1-70B-class models scaled across DP$\times$TP topologies (single-replica TP=2 and dual-replica DP=2$\times$TP=2) & 2--4$\times$H100 (TP=2; DP=1 / DP=2) \\
\addlinespace
W4 & Agentic bursts / programmatic serving & Tool-calling assistants (Cursor, Copilot, Claude Code), RAG pipelines with cite-check, reasoning chains (ToT / SoT), computer-use agents, back-end LLM enrichment pipelines & 1$\times$H100 \\
\addlinespace
W5 & Heterogeneous chat (no-regress)      & Default ChatGPT / Claude / Gemini chat, open-source hosted chat (HuggingChat, LMSYS Arena), general-purpose inference APIs without enforced templating & 1$\times$H100 \\
\bottomrule
\end{tabular}
\end{table}

\subsection{W1 Full-Lattice Heavy-Load Tables}
\label{app:w1_tables}

The three tables below report W1's full 7-policy lattice on heavy load: throughput, cache hit, mean TTFT, and mean E2E on SGLang (Table~\ref{tab:w1_sglang}) and vLLM (Table~\ref{tab:w1_vllm}); decode-side mean TPOT and ITL on both engines (Table~\ref{tab:w1_decode}). The body's W1 narrative (\S\ref{sec:w1}) summarizes these via the figures and ablation paragraphs.

\begin{table}[!htb]
\caption{W1 heavy load on \textbf{SGLang}: throughput, cache hit rate, mean TTFT, and mean E2E. Means across 3 seeds. PE = PEEK's queue-aware eviction (\texttt{demand\_cluster}); GM = group-major admission; DL = dynamic-lane fairness. \texttt{cLPM} alone is the cluster-LPM scheduler with stock LRU eviction (no group-major, no dynamic-lane).}
\label{tab:w1_sglang}
\centering
\scriptsize
\setlength{\tabcolsep}{2.5pt}
\begin{tabular}{llccccccc}
\toprule
\textbf{Cell} & \textbf{Metric} & \textbf{LPM+LRU} & \textbf{LPM+PE} & \textbf{cLPM} & \textbf{cLPM+GM} & \textbf{cLPM+GM+DL} & \textbf{cLPM+GM+PE} & \textbf{cLPM+GM+DL+PE} \\
\midrule
\multirow{4}{*}{A (2$\times$)}  & tput (req/s)        &  7.43 &  7.36 & 14.46 & 14.80 & 14.98 & 14.89 & 15.03 \\
                                & cache hit (\%)      & 65.8  & 65.4  & 93.7  & 93.6  & 93.5  & 93.5  & 93.6  \\
                                & TTFT (s)            & 42.6  & 43.1  & 11.0  & 10.8  & 10.4  & 10.5  & 10.3  \\
                                & E2E (s)             & 47.8  & 48.3  & 16.9  & 16.4  & 16.0  & 16.2  & 15.9  \\
\multirow{4}{*}{B (4$\times$)}  & tput (req/s)        &  6.73 &  6.70 & 13.74 & 14.34 & 14.49 & 14.66 & 14.55 \\
                                & cache hit (\%)      & 59.2  & 58.8  & 91.9  & 93.3  & 93.3  & 93.3  & 93.2  \\
                                & TTFT (s)            & 55.0  & 55.3  & 16.5  & 15.4  & 14.9  & 14.9  & 15.0  \\
                                & E2E (s)             & 60.3  & 60.8  & 22.5  & 21.1  & 20.8  & 20.8  & 20.8  \\
\multirow{4}{*}{C (8$\times$)}  & tput (req/s)        &  1.42 &  1.41 &  3.72 &  3.75 &  3.71 &  3.75 &  3.73 \\
                                & cache hit (\%)      & 38.3  & 38.3  & 88.3  & 89.4  & 89.2  & 89.2  & 88.8  \\
                                & TTFT (s)            & 233.2 & 234.0 &  36.9 &  32.7 &  31.5 &  32.6 &  29.7 \\
                                & E2E (s)             & 238.2 & 239.0 &  42.6 &  38.3 &  37.2 &  38.3 &  35.3 \\
\multirow{4}{*}{D (16$\times$)} & tput (req/s)        &  1.22 &  1.20 &  4.28 &  4.42 &  4.36 &  4.44 &  4.41 \\
                                & cache hit (\%)      & 30.8  & 30.9  & 91.0  & 92.2  & 92.2  & 92.4  & 92.1  \\
                                & TTFT (s)            & 335.5 & 340.2 &  62.5 &  59.6 &  60.3 &  59.9 &  59.7 \\
                                & E2E (s)             & 340.6 & 345.4 &  69.1 &  66.1 &  66.7 &  66.3 &  66.2 \\
\bottomrule
\end{tabular}
\end{table}

\begin{table}[!htb]
\caption{W1 heavy load on \textbf{vLLM}: throughput, cache hit rate, mean TTFT, and mean E2E. All vLLM cells run with Automatic Prefix Caching (APC) enabled. Means across 3 seeds. Column abbreviations: PE = PEEK's queue-aware eviction; GM = group-major admission; DL = dynamic-lane fairness.}
\label{tab:w1_vllm}
\centering
\scriptsize
\setlength{\tabcolsep}{2.5pt}
\begin{tabular}{llccccccc}
\toprule
\textbf{Cell} & \textbf{Metric} & \textbf{FCFS(APC)+LRU} & \textbf{FCFS(APC)+PE} & \textbf{cLPM} & \textbf{cLPM+GM} & \textbf{cLPM+GM+DL} & \textbf{cLPM+GM+PE} & \textbf{cLPM+GM+DL+PE} \\
\midrule
\multirow{4}{*}{A (2$\times$)}  & tput (req/s)        &  8.50 &  8.34 & 11.58 & 13.14 & 11.41 & 12.96 & 13.04 \\
                                & cache hit (\%)      & 66.3  & 66.1  & 87.5  & 86.4  & 86.7  & 86.8  & 86.7  \\
                                & TTFT (s)            & 14.0  & 14.4  &  6.0  &  4.2  &  6.1  &  4.3  &  4.2  \\
                                & E2E (s)             & 19.3  & 19.9  & 11.1  & 10.4  & 11.2  & 10.7  & 10.4  \\
\multirow{4}{*}{B (4$\times$)}  & tput (req/s)        &  7.52 &  7.31 & 10.83 & 11.76 & 10.65 & 11.56 & 11.57 \\
                                & cache hit (\%)      & 58.1  & 57.9  & 83.7  & 81.1  & 83.5  & 81.7  & 81.7  \\
                                & TTFT (s)            & 17.1  & 17.7  &  8.0  &  6.5  &  7.8  &  6.5  &  6.4  \\
                                & E2E (s)             & 22.4  & 23.1  & 13.2  & 12.6  & 13.1  & 12.8  & 12.5  \\
\multirow{4}{*}{C (8$\times$)}  & tput (req/s)        &  1.68 &  1.82 &  4.90 &  4.36 &  4.80 &  4.24 &  4.25 \\
                                & cache hit (\%)      & 37.7  & 38.0  & 86.5  & 86.2  & 86.2  & 85.8  & 86.3  \\
                                & TTFT (s)            & 97.8  & 86.6  & 16.9  & 21.8  & 18.5  & 22.6  & 22.4  \\
                                & E2E (s)             & 102.6 & 91.5  & 22.4  & 27.2  & 23.8  & 28.7  & 28.5  \\
\multirow{4}{*}{D (16$\times$)} & tput (req/s)        &  1.54 &  1.46 &  3.04 &  3.60 &  2.96 &  3.57 &  3.52 \\
                                & cache hit (\%)      & 32.8  & 31.1  & 81.8  & 78.9  & 80.9  & 79.2  & 79.0  \\
                                & TTFT (s)            & 106.3 & 111.1 & 39.4  & 35.0  & 40.5  & 34.4  & 33.2  \\
                                & E2E (s)             & 111.3 & 116.3 & 44.4  & 40.4  & 45.6  & 40.4  & 39.0  \\
\bottomrule
\end{tabular}
\end{table}

\begin{table}[!htb]
\caption{W1 heavy load decode-side metrics: mean TPOT and mean ITL on both engines. Means across 3 seeds. TPOT and ITL agree to within 0.1\,ms in every cell. All vLLM cells run with APC enabled. Column abbreviations: PE = PEEK's queue-aware eviction; GM = group-major admission; DL = dynamic-lane fairness.}
\label{tab:w1_decode}
\centering
\scriptsize
\setlength{\tabcolsep}{2.5pt}
\begin{tabular}{llccccccc}
\toprule
\multicolumn{9}{c}{\emph{SGLang}} \\
\midrule
\textbf{Cell} & \textbf{Metric} & \textbf{LPM+LRU} & \textbf{LPM+PE} & \textbf{cLPM} & \textbf{cLPM+GM} & \textbf{cLPM+GM+DL} & \textbf{cLPM+GM+PE} & \textbf{cLPM+GM+DL+PE} \\
\midrule
\multirow{2}{*}{A (2$\times$)}  & TPOT (ms) & 45.9 & 46.7 & 52.0 & 49.4 & 49.6 & 50.9 & 50.5 \\
                                & ITL  (ms) & 45.8 & 46.5 & 51.9 & 49.3 & 49.6 & 50.9 & 50.4 \\
\multirow{2}{*}{B (4$\times$)}  & TPOT (ms) & 46.9 & 47.6 & 52.4 & 50.2 & 51.4 & 52.1 & 50.5 \\
                                & ITL  (ms) & 46.9 & 47.7 & 52.3 & 50.2 & 51.3 & 52.0 & 50.5 \\
\multirow{2}{*}{C (8$\times$)}  & TPOT (ms) & 44.4 & 44.4 & 49.9 & 49.4 & 49.9 & 50.2 & 49.1 \\
                                & ITL  (ms) & 44.5 & 44.5 & 50.0 & 49.4 & 49.9 & 50.2 & 49.1 \\
\multirow{2}{*}{D (16$\times$)} & TPOT (ms) & 44.1 & 45.0 & 56.0 & 55.5 & 55.4 & 54.9 & 55.6 \\
                                & ITL  (ms) & 44.2 & 45.0 & 56.0 & 55.5 & 55.3 & 55.0 & 55.7 \\
\midrule
\multicolumn{9}{c}{\emph{vLLM}} \\
\midrule
\textbf{Cell} & \textbf{Metric} & \textbf{FCFS(APC)+LRU} & \textbf{FCFS(APC)+PE} & \textbf{cLPM} & \textbf{cLPM+GM} & \textbf{cLPM+GM+DL} & \textbf{cLPM+GM+PE} & \textbf{cLPM+GM+DL+PE} \\
\midrule
\multirow{2}{*}{A (2$\times$)}  & TPOT (ms) & 43.5 & 44.3 & 44.3 & 50.0 & 44.8 & 51.7 & 50.3 \\
                                & ITL  (ms) & 43.5 & 44.3 & 44.2 & 50.0 & 44.6 & 51.4 & 50.2 \\
\multirow{2}{*}{B (4$\times$)}  & TPOT (ms) & 42.2 & 43.2 & 44.8 & 49.4 & 45.1 & 51.1 & 49.8 \\
                                & ITL  (ms) & 42.2 & 43.2 & 44.5 & 49.4 & 45.0 & 51.0 & 49.8 \\
\multirow{2}{*}{C (8$\times$)}  & TPOT (ms) & 39.5 & 39.0 & 42.7 & 43.1 & 42.4 & 49.3 & 49.3 \\
                                & ITL  (ms) & 39.4 & 39.0 & 43.0 & 43.1 & 42.4 & 49.3 & 49.3 \\
\multirow{2}{*}{D (16$\times$)} & TPOT (ms) & 40.5 & 41.8 & 42.1 & 43.7 & 42.5 & 48.4 & 46.2 \\
                                & ITL  (ms) & 40.5 & 41.8 & 42.1 & 43.7 & 42.5 & 48.3 & 46.7 \\
\bottomrule
\end{tabular}
\end{table}

\subsection{W1 Throughput and E2E Figures}
\label{app:w1_figs}

These two figures complement the TTFT figure in the body (Figure~\ref{fig:w1_ttft}) and the full-lattice tables in Appendix~\ref{app:w1_tables}.

\begin{figure}[!htb]
\centering
\begin{subfigure}[t]{0.48\columnwidth}\centering\includegraphics[width=\linewidth]{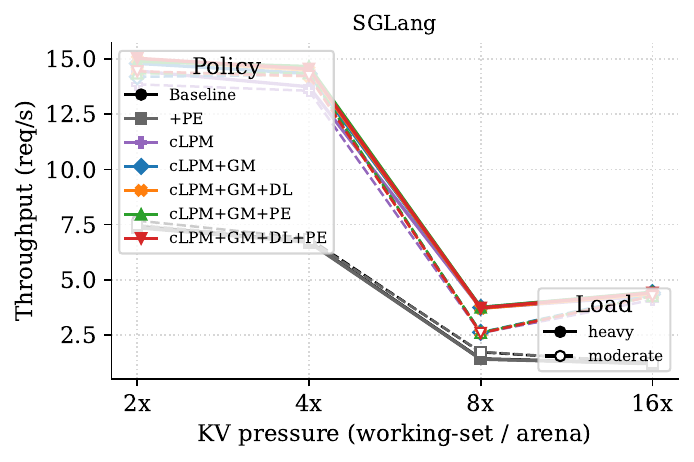}\caption{SGLang.}\label{fig:w1_sglang_rps}\end{subfigure}\hfill
\begin{subfigure}[t]{0.48\columnwidth}\centering\includegraphics[width=\linewidth]{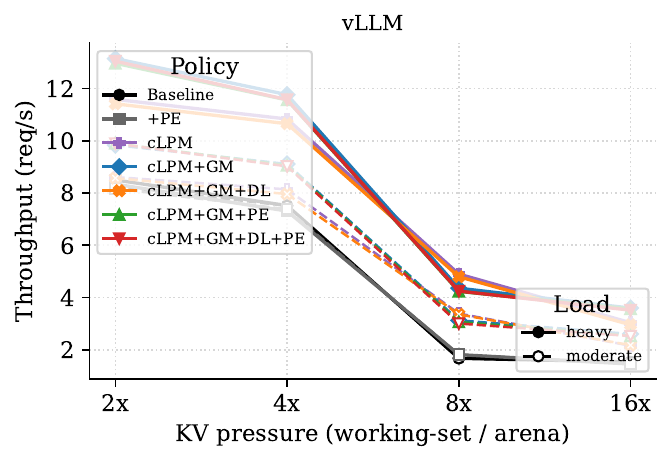}\caption{vLLM.}\label{fig:w1_vllm_rps}\end{subfigure}
\caption{W1 throughput vs.\ KV pressure. Each panel overlays both load levels: solid lines with filled markers = heavy, dashed lines with hollow markers = moderate.}
\label{fig:w1_rps}
\end{figure}

\begin{figure}[!htb]
\centering
\begin{subfigure}[t]{0.48\columnwidth}\centering\includegraphics[width=\linewidth]{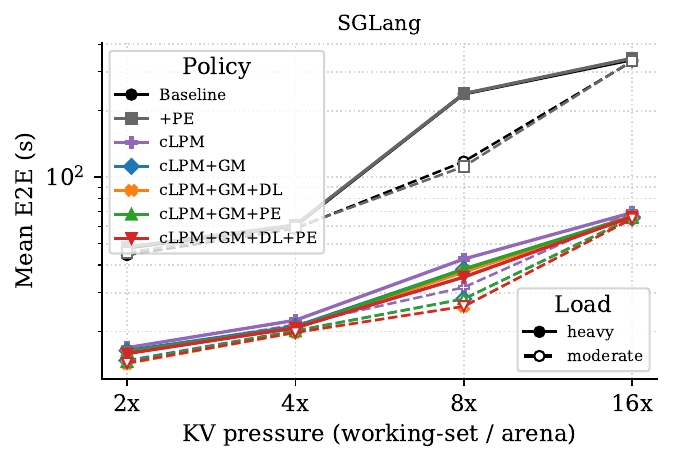}\caption{SGLang.}\label{fig:w1_sglang_e2e}\end{subfigure}\hfill
\begin{subfigure}[t]{0.48\columnwidth}\centering\includegraphics[width=\linewidth]{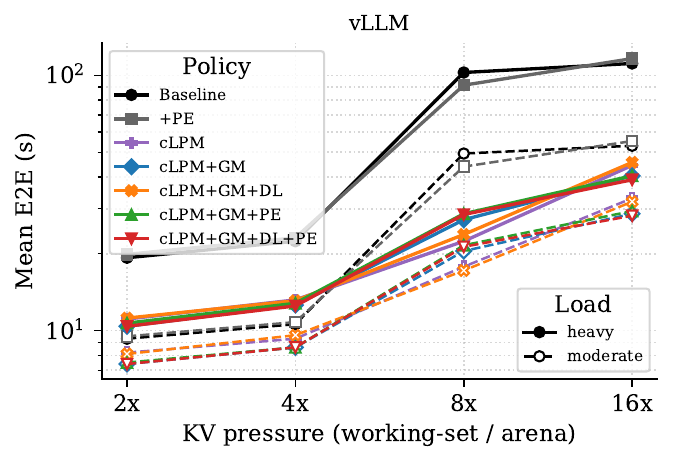}\caption{vLLM.}\label{fig:w1_vllm_e2e}\end{subfigure}
\caption{W1 mean E2E latency vs.\ KV pressure (log y-axis). Conventions as in Figure~\ref{fig:w1_ttft}. Same pattern: baselines climb, PEEK stays bounded.}
\label{fig:w1_e2e}
\end{figure}

\subsection{W2 Heavy and Moderate Load Tables}
\label{app:w2_tables}

\begin{table}[!htb]
\caption{W2 full configuration. Top: shared across all W2 runs. Bottom: per-engine, per-load offered arrival rate and request-count; both rates target sustained queue depths matching the paper-wide moderate/heavy definitions and run slightly above $r_{\text{sat}}$ (errored counts in single digits). The repository excerpts (Python files within shared repos) provide stable shared prefixes that exceed the model's KV arena at 8192 tokens, exercising eviction in steady state.}
\label{tab:w2_cells}
\centering
\footnotesize
\setlength{\tabcolsep}{4pt}

\begin{tabular}{p{0.22\linewidth} p{0.72\linewidth}}
\toprule
\multicolumn{2}{c}{\textbf{Shared across W2 (cell B, 7$\times$ KV pressure)}} \\
\midrule
Hardware             & 1$\times$NVIDIA H100 80\,GB (bf16) \\
Model                & Qwen2.5-32B-Instruct \\
Engines              & SGLang 0.5.9 (B1=LPM+LRU baseline); vLLM 0.19.1 (B2=FCFS+APC+LRU baseline) \\
Dataset              & RepoBench-Python v1.1 (\texttt{cross\_file\_first} split): $G$=40 repository contexts, each truncated to a 8192-token shared prefix; per-request continuation question appended after the shared prefix \\
Group sampling       & Zipf-$\alpha$=1.0 over the $G$=40 prefixes \\
Arrival process      & Poisson (inter-arrival $\sim$Exp($\lambda$)) \\
Decode               & Variable, sampled from the mix 10:128, 25:512, 30:1024, 25:2048, 10:4096 (mean $\approx$1460 tokens per request) \\
Concurrency cap      & 256 (both engines) \\
Seeds                & 42, 142, 242 (3 replicates per cell; reported numbers are means across seeds) \\
Warmup               & First 100 requests excluded from metrics \\
Memory budget        & SGLang \texttt{mem\_fraction\_static}=0.88; vLLM \texttt{gpu\_memory\_utilization}=0.9 \\
Policies             & B1 (SGLang) / B2 (vLLM); \textbf{cLPM+GM+DL+PE} (full PEEK co-design) \\
\bottomrule
\end{tabular}

\vspace{0.5em}
\begin{tabular}{lcccccc}
\toprule
& \multicolumn{3}{c}{\textbf{SGLang}} & \multicolumn{3}{c}{\textbf{vLLM}} \\
\cmidrule(lr){2-4}\cmidrule(lr){5-7}
\textbf{Load} & \textbf{rate (req/s)} & \textbf{$N$} & \textbf{warmup} & \textbf{rate (req/s)} & \textbf{$N$} & \textbf{warmup} \\
\midrule
moderate      & 0.40 & 1000 & 100 & 0.15 & 500 & 100 \\
heavy         & 0.45 & 1000 & 100 & 0.20 & 500 & 100 \\
\bottomrule
\end{tabular}
\end{table}

\begin{table}[!htb]
\caption{W2 on \textbf{SGLang} (cell B, 7$\times$ KV pressure): throughput, cache hit rate, mean TTFT, mean TPOT, mean E2E. Means across 3 seeds. Both rates run slightly above $r_{\text{sat}}$ (errored counts in the single digits), so throughput represents a saturated regime; cache-hit and TTFT deltas remain interpretable.}
\label{tab:w2_sglang}
\centering
\footnotesize
\setlength{\tabcolsep}{4pt}
\begin{tabular}{llcc}
\toprule
\textbf{Load} & \textbf{Metric} & \textbf{LPM+LRU} & \textbf{cLPM+GM+DL+PE} \\
\midrule
\multirow{5}{*}{moderate} & tput (req/s)   & 0.289 & 0.330 \\
                          & cache hit (\%) & 62.4  & 79.4  \\
                          & TTFT (s)       & 391.1 & 178.8 \\
                          & TPOT (ms)      & 32.6  & 31.8  \\
                          & E2E (s)        & 419.2 & 206.8 \\
\multirow{5}{*}{heavy}    & tput (req/s)   & 0.283 & 0.357 \\
                          & cache hit (\%) & 60.0  & 83.2  \\
                          & TTFT (s)       & 543.4 & 212.2 \\
                          & TPOT (ms)      & 31.9  & 33.4  \\
                          & E2E (s)        & 571.5 & 241.6 \\
\bottomrule
\end{tabular}
\end{table}

\begin{table}[!htb]
\caption{W2 on \textbf{vLLM} (cell B, 7$\times$ KV pressure): throughput, cache hit rate, mean TTFT, mean TPOT, mean E2E. All cells run with APC enabled. Means across 3 seeds. errored$=$0 across all cells.}
\label{tab:w2_vllm}
\centering
\footnotesize
\setlength{\tabcolsep}{4pt}
\begin{tabular}{llcc}
\toprule
\textbf{Load} & \textbf{Metric} & \textbf{FCFS(APC)+LRU} & \textbf{cLPM+GM+DL+PE} \\
\midrule
\multirow{5}{*}{moderate} & tput (req/s)   & 0.114 & 0.114 \\
                          & cache hit (\%) & 29.4  & 33.0  \\
                          & TTFT (s)       & 28.4  & 8.3   \\
                          & TPOT (ms)      & 27.4  & 27.8  \\
                          & E2E (s)        & 51.4  & 31.6  \\
\multirow{5}{*}{heavy}    & tput (req/s)   & 0.143 & 0.151 \\
                          & cache hit (\%) & 29.5  & 40.1  \\
                          & TTFT (s)       & 90.7  & 24.7  \\
                          & TPOT (ms)      & 28.8  & 31.2  \\
                          & E2E (s)        & 114.6 & 50.5  \\
\bottomrule
\end{tabular}
\end{table}

\subsection{W2 Throughput and E2E Figures}
\label{app:w2_figs}

\begin{figure}[!htb]
\centering
\begin{subfigure}[t]{0.48\columnwidth}\centering\includegraphics[width=\linewidth]{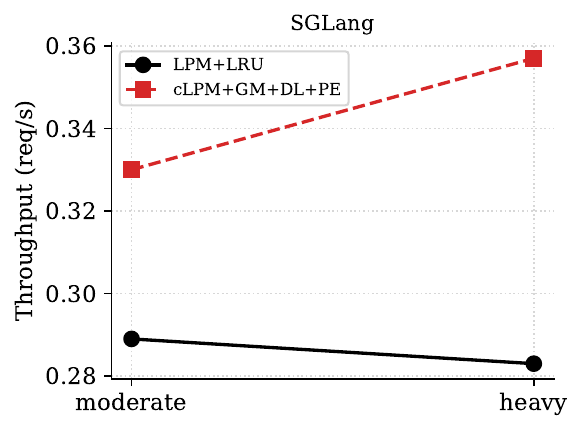}\caption{SGLang.}\label{fig:w2_sglang_rps}\end{subfigure}\hfill
\begin{subfigure}[t]{0.48\columnwidth}\centering\includegraphics[width=\linewidth]{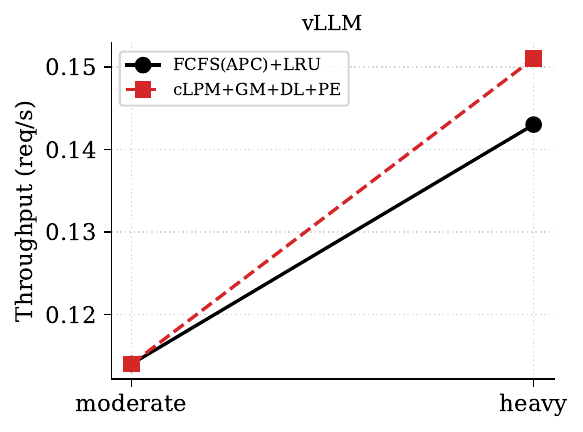}\caption{vLLM.}\label{fig:w2_vllm_rps}\end{subfigure}
\caption{W2 throughput on cell B, both engines and both load levels.}
\label{fig:w2_rps}
\end{figure}

\begin{figure}[!htb]
\centering
\begin{subfigure}[t]{0.48\columnwidth}\centering\includegraphics[width=\linewidth]{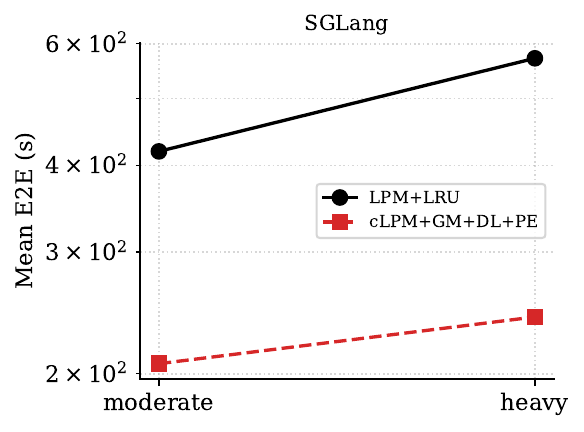}\caption{SGLang.}\label{fig:w2_sglang_e2e}\end{subfigure}\hfill
\begin{subfigure}[t]{0.48\columnwidth}\centering\includegraphics[width=\linewidth]{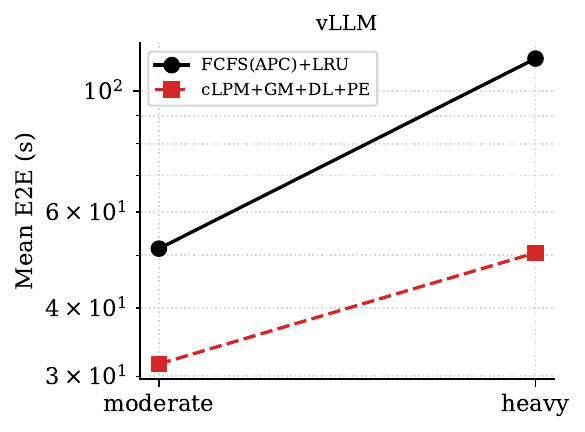}\caption{vLLM.}\label{fig:w2_vllm_e2e}\end{subfigure}
\caption{W2 mean E2E latency on cell B (log y-axis), both engines and both load levels.}
\label{fig:w2_e2e}
\end{figure}

\subsection{W3 Heavy-Load Tables (DP=1 and DP=2)}
\label{app:w3_tables}

\begin{table}[!htb]
\caption{W3 heavy load on \textbf{SGLang} (Llama-3.1-70B, TP=2): throughput, cache hit, mean TTFT, mean TPOT, mean E2E at DP=1 (2$\times$H100) and DP=2 (4$\times$H100). Means across 3 seeds.}
\label{tab:w3_sglang}
\centering
\footnotesize
\setlength{\tabcolsep}{4pt}
\begin{tabular}{llcccc}
\toprule
& & \multicolumn{2}{c}{\textbf{DP=1 (2$\times$H100)}} & \multicolumn{2}{c}{\textbf{DP=2 (4$\times$H100)}} \\
\cmidrule(lr){3-4} \cmidrule(lr){5-6}
\textbf{Cell} & \textbf{Metric} & \textbf{LPM+LRU} & \textbf{cLPM+GM+DL+PE} & \textbf{LPM+LRU} & \textbf{cLPM+GM+DL+PE} \\
\midrule
\multirow{5}{*}{B (RAG-like)}  & tput (req/s)   & 0.464 & 1.237 & 1.043  & 1.689  \\
                                & cache hit (\%) & 55.1  & 98.2  & 87.8   & 96.9   \\
                                & TTFT (s)       & 324.2 & 85.3  & 180.7  & 91.4   \\
                                & TPOT (ms)      & 32.4  & 32.8  & 31.9   & 32.8   \\
                                & E2E (s)        & 330.1 & 91.3  & 192.6  & 103.6  \\
\multirow{5}{*}{C (chat-like)} & tput (req/s)   & 2.896 & 6.022 & 7.757  & 13.634 \\
                                & cache hit (\%) & 44.6  & 84.2  & 72.9   & 91.5   \\
                                & TTFT (s)       & 46.6  & 15.7  & 28.7   & 10.2   \\
                                & TPOT (ms)      & 47.6  & 44.3  & 37.0   & 36.7   \\
                                & E2E (s)        & 50.0  & 18.9  & 33.4   & 14.9   \\
\bottomrule
\end{tabular}
\end{table}

\begin{table}[!htb]
\caption{W3 heavy load on \textbf{vLLM} (Llama-3.1-70B, TP=2; APC enabled): throughput, cache hit, mean TTFT, mean TPOT, mean E2E at DP=1 and DP=2. Means across 3 seeds.}
\label{tab:w3_vllm}
\centering
\footnotesize
\setlength{\tabcolsep}{4pt}
\begin{tabular}{llcccc}
\toprule
& & \multicolumn{2}{c}{\textbf{DP=1 (2$\times$H100)}} & \multicolumn{2}{c}{\textbf{DP=2 (4$\times$H100)}} \\
\cmidrule(lr){3-4} \cmidrule(lr){5-6}
\textbf{Cell} & \textbf{Metric} & \textbf{FCFS(APC)+LRU} & \textbf{cLPM+GM+DL+PE} & \textbf{FCFS(APC)+LRU} & \textbf{cLPM+GM+DL+PE} \\
\midrule
\multirow{5}{*}{B (RAG-like)}  & tput (req/s)   & 0.241 & 1.072 & 0.478  & 2.152  \\
                                & cache hit (\%) & 37.7  & 97.1  & 55.4   & 97.3   \\
                                & TTFT (s)       & 559.6 & 83.2  & 384.4  & 54.3   \\
                                & TPOT (ms)      & 36.0  & 65.9  & 42.1   & 93.9   \\
                                & E2E (s)        & 572.7 & 103.4 & 399.6  & 83.8   \\
\multirow{5}{*}{C (chat-like)} & tput (req/s)   & 1.798 & 3.728 & 4.040  & 11.440 \\
                                & cache hit (\%) & 37.2  & 76.5  & 51.4   & 92.7   \\
                                & TTFT (s)       & 75.8  & 24.9  & 60.5   & 12.3   \\
                                & TPOT (ms)      & 32.1  & 48.1  & 32.8   & 43.7   \\
                                & E2E (s)        & 79.9  & 31.0  & 64.7   & 17.9   \\
\bottomrule
\end{tabular}
\end{table}

\subsection{W3 Throughput and E2E Figures (DP=1 and DP=2)}
\label{app:w3_figs}

\begin{figure}[!htb]
\centering
\begin{subfigure}[t]{0.48\columnwidth}\centering\includegraphics[width=\linewidth]{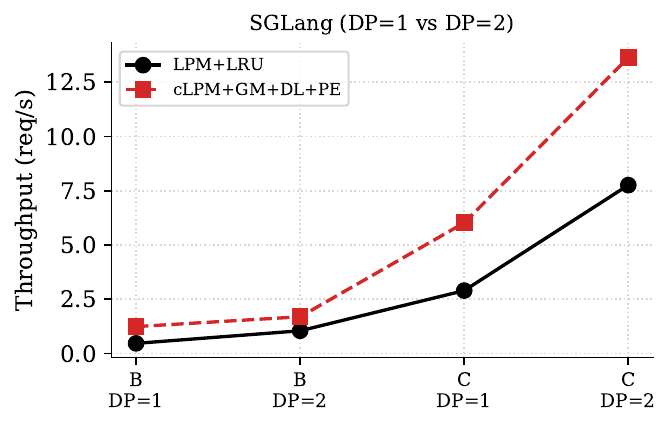}\caption{SGLang.}\label{fig:w3_sglang_rps}\end{subfigure}\hfill
\begin{subfigure}[t]{0.48\columnwidth}\centering\includegraphics[width=\linewidth]{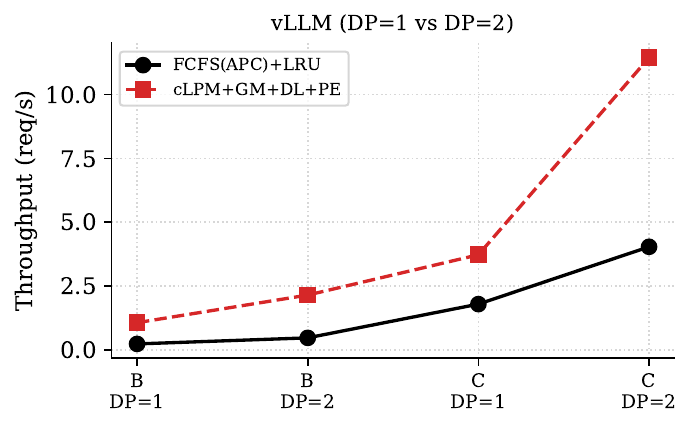}\caption{vLLM.}\label{fig:w3_vllm_rps}\end{subfigure}
\caption{W3 throughput, DP=1 vs DP=2, heavy load.}
\label{fig:w3_rps}
\end{figure}

\begin{figure}[!htb]
\centering
\begin{subfigure}[t]{0.48\columnwidth}\centering\includegraphics[width=\linewidth]{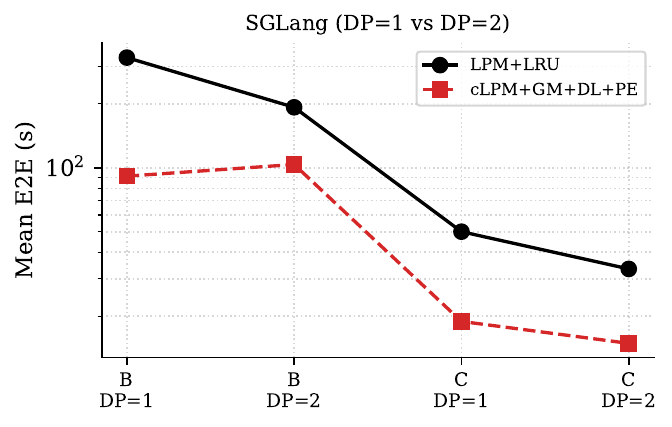}\caption{SGLang.}\label{fig:w3_sglang_e2e}\end{subfigure}\hfill
\begin{subfigure}[t]{0.48\columnwidth}\centering\includegraphics[width=\linewidth]{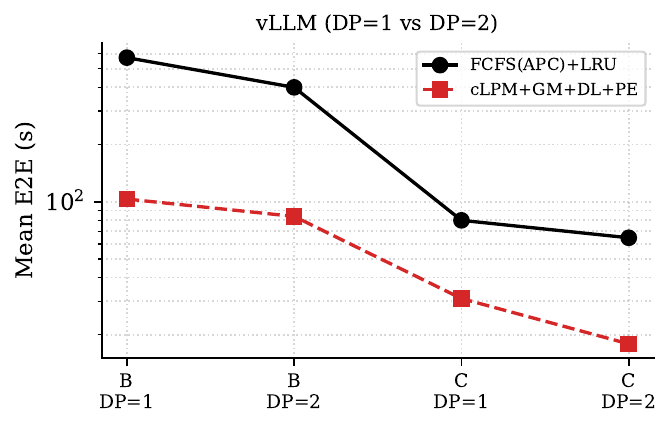}\caption{vLLM.}\label{fig:w3_vllm_e2e}\end{subfigure}
\caption{W3 mean E2E latency (log y-axis), DP=1 vs DP=2, heavy load.}
\label{fig:w3_e2e}
\end{figure}

\subsection{W4 Heavy-Load Tables}
\label{app:w4_tables}

\begin{table}[!htb]
\caption{W4 heavy load on \textbf{SGLang}: throughput, cache hit rate, mean TTFT, mean TPOT, mean E2E. Means across 3 seeds.}
\label{tab:w4_sglang}
\centering
\footnotesize
\setlength{\tabcolsep}{4pt}
\begin{tabular}{llccc}
\toprule
\textbf{Scenario} & \textbf{Metric} & \textbf{FCFS+LRU} & \textbf{LPM+LRU} & \textbf{cLPM+GM+DL+PE} \\
\midrule
\multirow{5}{*}{agentic\_only}   & tput (req/s)   & 3.58 & 3.58 & 3.62 \\
                                  & cache hit (\%) & 59.5 & 60.4 & 60.7 \\
                                  & TTFT (ms)      & 1175 & 1129 & 1185 \\
                                  & TPOT (ms)      & 65.8 & 63.9 & 63.7 \\
                                  & E2E (s)        & 22.4 & 22.0 & 21.8 \\
\multirow{5}{*}{agentic\_shared} & tput (req/s)   & 3.59 & 3.58 & 3.62 \\
                                  & cache hit (\%) & 59.7 & 60.0 & 61.0 \\
                                  & TTFT (ms)      & 1181 & 1142 & 1157 \\
                                  & TPOT (ms)      & 66.3 & 65.1 & 62.8 \\
                                  & E2E (s)        & 22.4 & 22.2 & 21.6 \\
\bottomrule
\end{tabular}
\end{table}

\begin{table}[!htb]
\caption{W4 heavy load on \textbf{vLLM}: throughput, cache hit rate, mean TTFT, mean TPOT, mean E2E. All cells run with APC enabled. Means across 3 seeds.}
\label{tab:w4_vllm}
\centering
\footnotesize
\setlength{\tabcolsep}{4pt}
\begin{tabular}{llcc}
\toprule
\textbf{Scenario} & \textbf{Metric} & \textbf{FCFS(APC)+LRU} & \textbf{cLPM+GM+DL+PE} \\
\midrule
\multirow{5}{*}{agentic\_only}   & tput (req/s)   & 3.57 & 3.58 \\
                                  & cache hit (\%) & 85.6 & 87.2 \\
                                  & TTFT (ms)      & 178  & 167  \\
                                  & TPOT (ms)      & 22.8 & 22.5 \\
                                  & E2E (s)        & 5.40 & 5.31 \\
\multirow{5}{*}{agentic\_shared} & tput (req/s)   & 3.54 & 3.54 \\
                                  & cache hit (\%) & 91.5 & 91.5 \\
                                  & TTFT (ms)      & 171  & 173  \\
                                  & TPOT (ms)      & 22.8 & 22.9 \\
                                  & E2E (s)        & 4.78 & 4.79 \\
\bottomrule
\end{tabular}
\end{table}

\subsection{W4 Figures}
\label{app:w4_figs}

\begin{figure}[!htb]
\centering
\begin{subfigure}[t]{0.48\columnwidth}\centering\includegraphics[width=\linewidth]{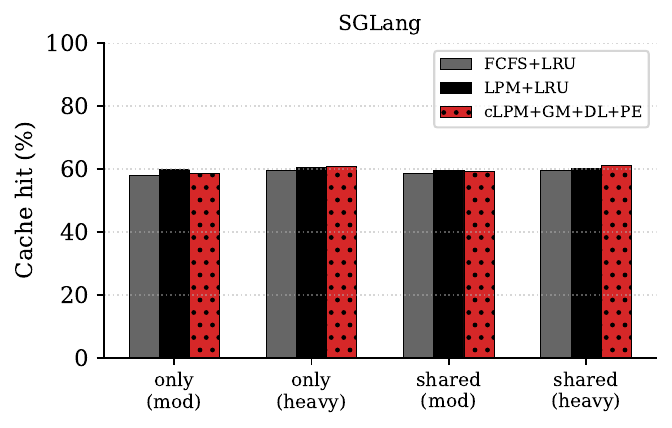}\caption{SGLang.}\label{fig:w4_sglang_cache_hit}\end{subfigure}\hfill
\begin{subfigure}[t]{0.48\columnwidth}\centering\includegraphics[width=\linewidth]{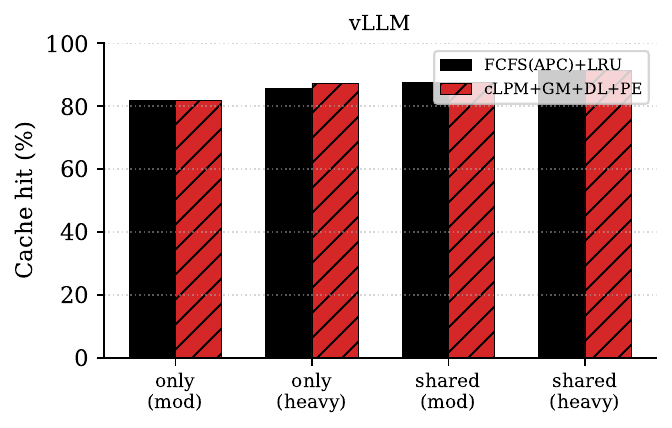}\caption{vLLM.}\label{fig:w4_vllm_cache_hit}\end{subfigure}
\caption{W4 cache hit rate by scenario and load level, side-by-side per engine.}
\label{fig:w4_cache_hit}
\end{figure}

\begin{figure}[!htb]
\centering
\begin{subfigure}[t]{0.48\columnwidth}\centering\includegraphics[width=\linewidth]{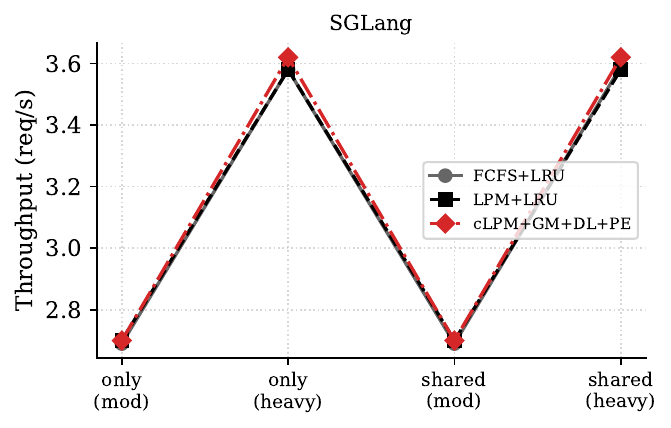}\caption{SGLang.}\label{fig:w4_sglang_rps}\end{subfigure}\hfill
\begin{subfigure}[t]{0.48\columnwidth}\centering\includegraphics[width=\linewidth]{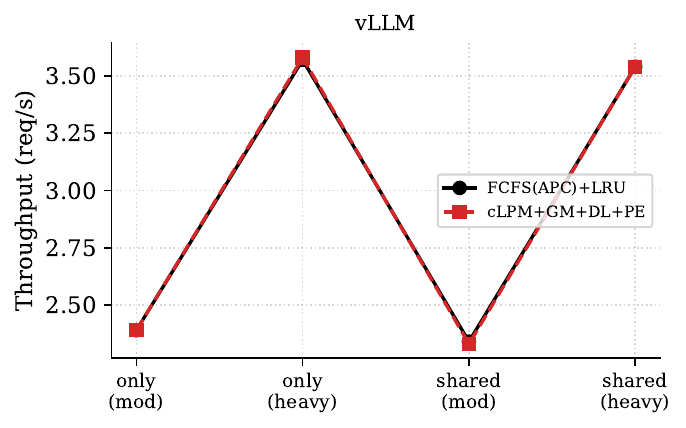}\caption{vLLM.}\label{fig:w4_vllm_rps}\end{subfigure}
\caption{W4 throughput by scenario and load level, side-by-side per engine.}
\label{fig:w4_rps}
\end{figure}

\begin{figure}[!htb]
\centering
\begin{subfigure}[t]{0.48\columnwidth}\centering\includegraphics[width=\linewidth]{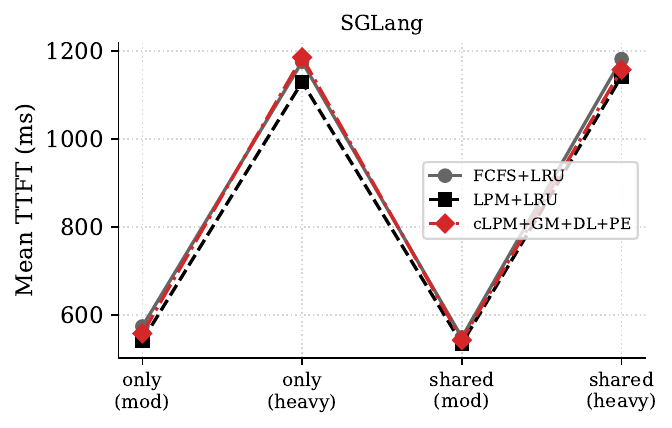}\caption{SGLang.}\label{fig:w4_sglang_ttft}\end{subfigure}\hfill
\begin{subfigure}[t]{0.48\columnwidth}\centering\includegraphics[width=\linewidth]{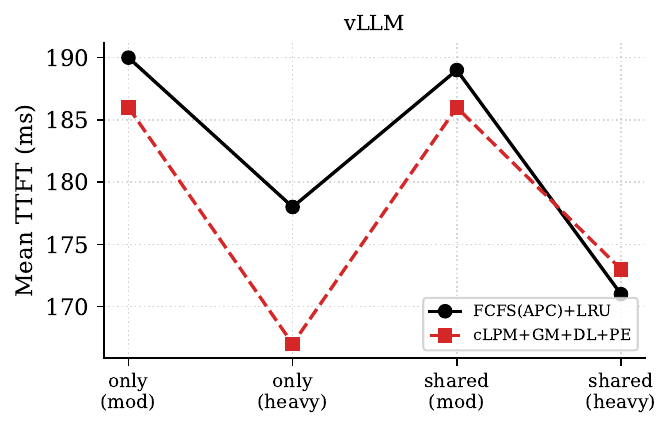}\caption{vLLM.}\label{fig:w4_vllm_ttft}\end{subfigure}
\caption{W4 mean TTFT by scenario and load level, side-by-side per engine.}
\label{fig:w4_ttft}
\end{figure}

\begin{figure}[!htb]
\centering
\begin{subfigure}[t]{0.48\columnwidth}\centering\includegraphics[width=\linewidth]{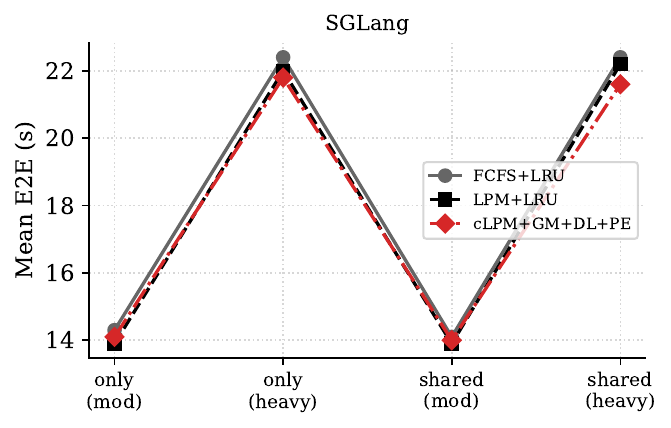}\caption{SGLang.}\label{fig:w4_sglang_e2e}\end{subfigure}\hfill
\begin{subfigure}[t]{0.48\columnwidth}\centering\includegraphics[width=\linewidth]{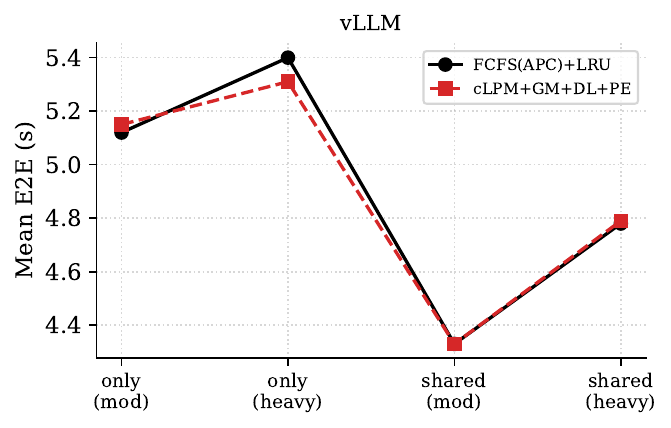}\caption{vLLM.}\label{fig:w4_vllm_e2e}\end{subfigure}
\caption{W4 mean E2E latency by scenario and load level, side-by-side per engine.}
\label{fig:w4_e2e}
\end{figure}

\subsection{W5 Heavy-Load Tables}
\label{app:w5_tables}

\begin{table}[!htb]
\caption{W5 heavy load on \textbf{SGLang}: throughput, cache hit, mean TTFT, mean TPOT, mean E2E. Means across available seeds (3 seeds for cell\_C\_short; cell\_C\_long: seed 242 only). Cache hit computed from Prometheus snapshots (RadixAttention requires exact-prefix matches; with no shared system prompt, baseline cache hit is near zero).}
\label{tab:w5_sglang}
\centering
\footnotesize
\setlength{\tabcolsep}{4pt}
\begin{tabular}{llcc}
\toprule
\textbf{Cell} & \textbf{Metric} & \textbf{LPM+LRU} & \textbf{cLPM+GM+DL+PE} \\
\midrule
\multirow{5}{*}{C\_long}  & tput (req/s)   & 7.83  & 8.20  \\
                          & cache hit (\%) & 0.7   & 10.2  \\
                          & TTFT (s)       & 21.7  & 19.4  \\
                          & TPOT (ms)      & 52.2  & 52.7  \\
                          & E2E (s)        & 28.3  & 26.0  \\
\multirow{5}{*}{C\_short} & tput (req/s)   & 21.89 & 21.68 \\
                          & cache hit (\%) & 3.7   & 3.7   \\
                          & TTFT (s)       & 3.6   & 3.6   \\
                          & TPOT (ms)      & 54.6  & 54.9  \\
                          & E2E (s)        & 10.05 & 10.14 \\
\bottomrule
\end{tabular}
\end{table}

\begin{table}[!htb]
\caption{W5 on \textbf{vLLM}: throughput, cache hit, mean TTFT, mean TPOT, mean E2E. Means across 3 seeds. APC's block-hash matching gives high reported cache hit on byte-recurring tokens across heterogeneous prompts (a property of the cache layer, not of structural sharing in the workload).}
\label{tab:w5_vllm}
\centering
\footnotesize
\setlength{\tabcolsep}{4pt}
\begin{tabular}{llcc}
\toprule
\textbf{Cell} & \textbf{Metric} & \textbf{FCFS(APC)+LRU} & \textbf{cLPM+GM+DL+PE} \\
\midrule
\multirow{5}{*}{C\_long}  & tput (req/s)   & 18.33 & 22.27 \\
                          & cache hit (\%) & 97.5  & 98.8  \\
                          & TTFT (s)       & 4.25  & 1.61  \\
                          & TPOT (ms)      & 73.6  & 74.3  \\
                          & E2E (s)        & 12.45 & 9.98  \\
\multirow{5}{*}{C\_short} & tput (req/s)   & 39.63 & 39.70 \\
                          & cache hit (\%) & 96.7  & 96.7  \\
                          & TTFT (s)       & 0.39  & 0.35  \\
                          & TPOT (ms)      & 43.9  & 44.0  \\
                          & E2E (s)        & 5.69  & 5.66  \\
\bottomrule
\end{tabular}
\end{table}

\subsection{W5 Figures}
\label{app:w5_figs}

\begin{figure}[!htb]
\centering
\begin{subfigure}[t]{0.48\columnwidth}\centering\includegraphics[width=\linewidth]{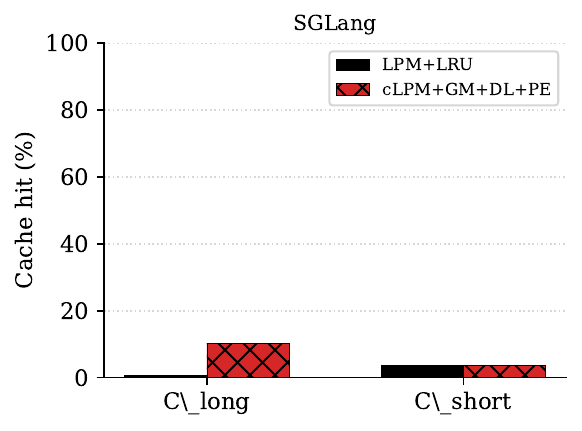}\caption{SGLang.}\label{fig:w5_sglang_cache_hit}\end{subfigure}\hfill
\begin{subfigure}[t]{0.48\columnwidth}\centering\includegraphics[width=\linewidth]{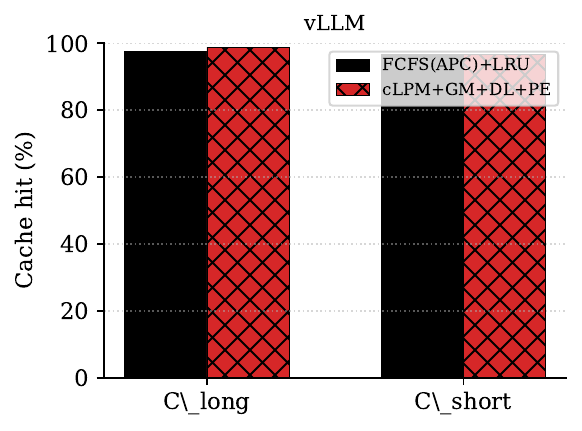}\caption{vLLM.}\label{fig:w5_vllm_cache_hit}\end{subfigure}
\caption{W5 cache hit rate by cell, both engines.}
\label{fig:w5_cache_hit}
\end{figure}

\begin{figure}[!htb]
\centering
\begin{subfigure}[t]{0.48\columnwidth}\centering\includegraphics[width=\linewidth]{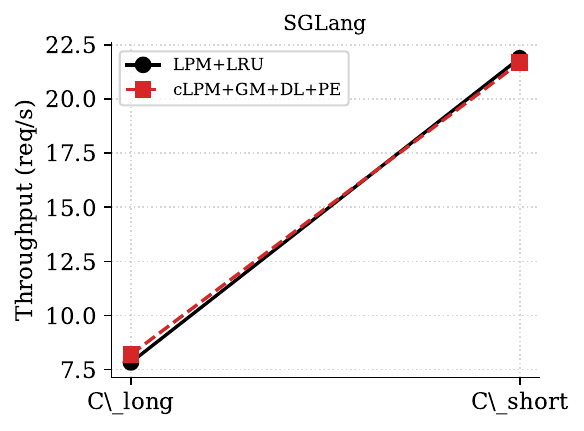}\caption{SGLang.}\label{fig:w5_sglang_rps}\end{subfigure}\hfill
\begin{subfigure}[t]{0.48\columnwidth}\centering\includegraphics[width=\linewidth]{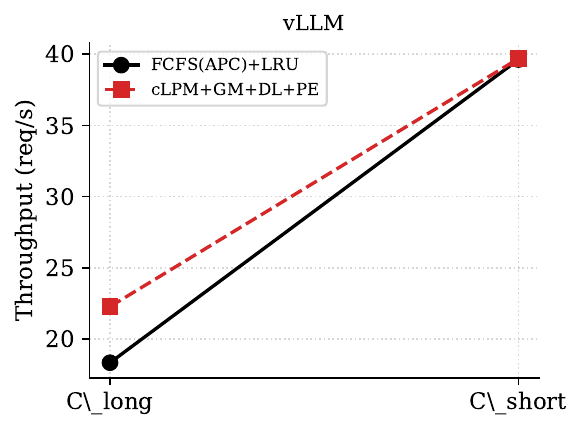}\caption{vLLM.}\label{fig:w5_vllm_rps}\end{subfigure}
\caption{W5 throughput by cell, both engines.}
\label{fig:w5_rps}
\end{figure}

\begin{figure}[!htb]
\centering
\begin{subfigure}[t]{0.48\columnwidth}\centering\includegraphics[width=\linewidth]{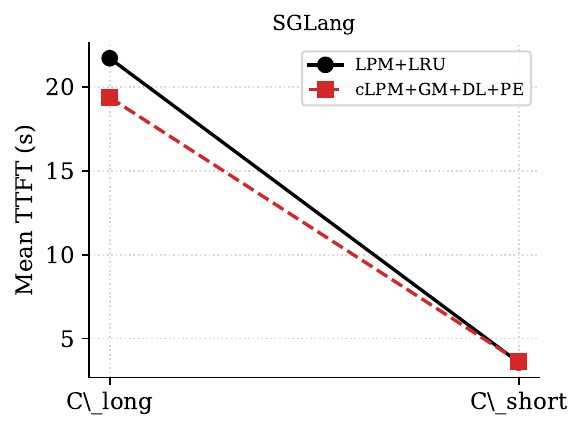}\caption{SGLang.}\label{fig:w5_sglang_ttft}\end{subfigure}\hfill
\begin{subfigure}[t]{0.48\columnwidth}\centering\includegraphics[width=\linewidth]{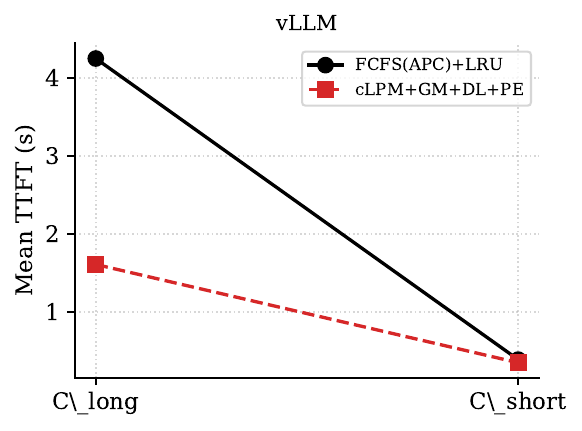}\caption{vLLM.}\label{fig:w5_vllm_ttft}\end{subfigure}
\caption{W5 mean TTFT by cell, both engines.}
\label{fig:w5_ttft}
\end{figure}

\begin{figure}[!htb]
\centering
\begin{subfigure}[t]{0.48\columnwidth}\centering\includegraphics[width=\linewidth]{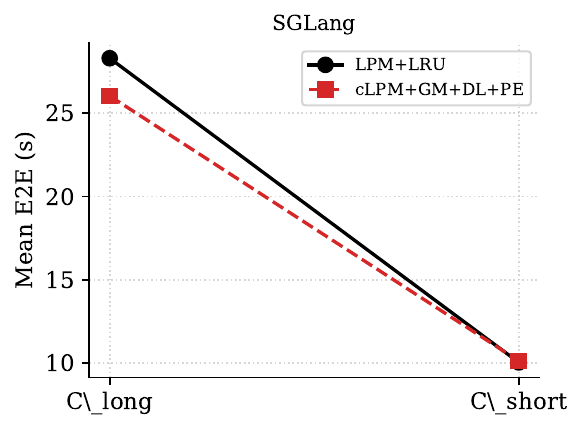}\caption{SGLang.}\label{fig:w5_sglang_e2e}\end{subfigure}\hfill
\begin{subfigure}[t]{0.48\columnwidth}\centering\includegraphics[width=\linewidth]{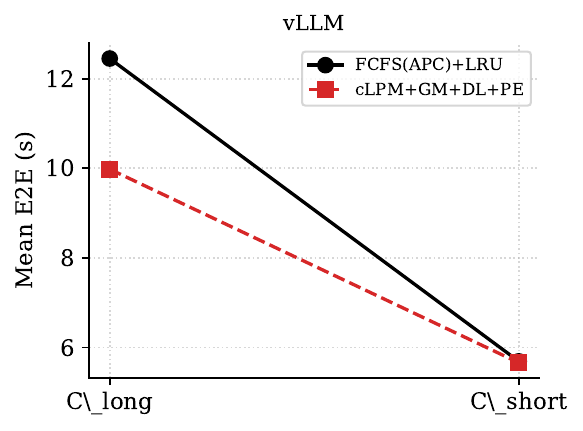}\caption{vLLM.}\label{fig:w5_vllm_e2e}\end{subfigure}
\caption{W5 mean E2E latency by cell, both engines.}
\label{fig:w5_e2e}
\end{figure}

\section{PEEK-offline}
\label{app:peek_offline}

\subsection{Mechanisms}
\label{app:offline_method}

In the offline (batch) regime the entire workload is known in advance and throughput is the sole objective; fairness and per-request latency tails are not first-order constraints. PEEK-offline exploits this by replacing the per-cycle pending tree of PEEK-online with a one-shot, full-batch construction, executed entirely as client-side preprocessing through the engine's public API.

\paragraph{Phase 1: Prefix trie DFS reordering.} PEEK-offline builds a token-level prefix trie over all $N$ submitted requests by inserting the first $D$ tokens of each prompt (default $D{=}128$); each node records the requests terminating there and the subtree request count. A depth-first traversal that visits children in descending subtree-count order emits a permutation $\pi$ in which prefix-sharing requests are adjacent, even when the engine cache is cold. Construction is $O(N \cdot D)$ and traversal is $O(|T|)$ with $|T| \ll N \cdot D$ under sharing; at $N{=}5{,}000$ the entire reorder completes in $<$50\,ms on a single CPU core.

\paragraph{Phase 2: Shadow LRU cache and wave-aware refinement.} The Phase~1 ordering is partitioned into \emph{waves} of size $W$ approximating the engine's running batch. A client-side shadow LRU cache, hashed identically to vLLM/SGLang ($k_i = \texttt{hash}(k_{i-1}, \texttt{tuple}(s[iB{:}(i{+}1)B]))$, $B{=}16$), simulates per-block residency in $O(1)$ amortized time and is reconciled against the engine's reported \texttt{cached\_tokens} after every completion. A second cache-aware DFS within each wave scores children by
\[
\texttt{key}(v) = \texttt{cache\_frac}(v)\!\cdot\!10^4 \;-\; \texttt{future\_refs}(v) \;+\; \texttt{count}(v)\!\cdot\!10^{-3},
\]
strictly preferring already-cached groups (exploit), deferring groups demanded by later waves (preserve), and using subtree count only as a tiebreaker. A queue-aware eviction hook scans the remaining batch to derive per-block reference counts, evicting unreferenced blocks first and protecting referenced blocks via log-dampened depth-weighted priority to avoid ossification.

\paragraph{Guard conditions.} Three cheap pre-checks skip all PEEK work when reordering cannot pay for itself: \emph{duplication} ($>$50\% exact-duplicate sequences), \emph{coverage} ($<$10\% of requests share $\geq$32 tokens), and \emph{depth} (average sharing depth $<$64 tokens). When any guard fires, PEEK-offline reduces to FCFS at $O(1)$ overhead.

\subsection{Experiment Results}
\label{app:offline_experiments}

PEEK-offline is evaluated on two configurations: \emph{Single-GPU} (Qwen2.5-14B-Instruct, 1$\times$H100 80\,GB, FP16) and \emph{Multi-GPU} (Qwen2.5-72B-Instruct, 2$\times$H100, TP=2, FP8), both on SGLang v0.5.9 and vLLM v0.10.x at production-default memory (SGLang \texttt{mem\_fraction\_static}=0.80, vLLM \texttt{gpu\_memory\_utilization}=0.90). Single-GPU uses six workloads (three high-sharing: shared system prompts on LooGLE, mixed traffic, multi-turn ShareGPT chat; three low-sharing: few-shot MMLU, code completion on HumanEval/MBPP, single-turn diverse). Multi-GPU uses six production scenarios (Legal RAG, Long-Doc Summarization, Multi-Agent, Enterprise Search, Mixed Traffic, plus a non-sharing control). Baselines are FCFS, LPM, and DFS-Weight with LRU/LFU on SGLang; FCFS and APC with LRU on vLLM.

\paragraph{Single-GPU.} On the three high-sharing workloads, PEEK-offline beats the strongest baseline by \textbf{1.1--1.6$\times$} on SGLang (cache hits 72--84\%) and \textbf{1.0--1.8$\times$} on vLLM (71--85\%). The largest gains are on shared system prompts where 100 prefix groups arrive randomly interleaved (FCFS achieves only 1\% hits); the smallest are on multi-turn chat, where sequential conversation turns supply natural adjacency (FCFS already at 50\% hits). Prefill reduction accounts for 89--97\% of total latency improvement; the remainder is decode-side, an indirect benefit of freeing GPU compute sooner. On the three low-sharing workloads, all policies are within 3\% of each other---the guards correctly skip reordering. P99/Mean latency ratios stay within 0.05 of the best baseline across all six workloads, confirming no starvation.

\paragraph{Multi-GPU.} Scaling to Qwen2.5-72B on 2$\times$H100, PEEK-offline beats the strongest baseline by \textbf{1.6--2.4$\times$} on SGLang (cache hits 90--96\%) across five high-sharing scenarios and by \textbf{1.6--8.0$\times$} on vLLM (90--98\%) across four scenarios. The largest gain is on Multi-Agent (16K shared prefixes), the smallest on Mixed Traffic (moderate natural adjacency). Non-sharing controls remain within 5\% of baseline.

\paragraph{Component ablation and scaling.} A reorder-only variant (DFS without shadow cache or eviction) matches the full system on every workload, while eviction-only delivers no benefit in isolation---DFS reordering supplies the temporal locality that queue-aware eviction then preserves. Sweeps confirm gains scale monotonically with cache pressure: prefix length 512$\to$4096 grows the gain from 1.0$\times$ to \textbf{3.0--3.8$\times$}; group count 20$\to$200 from $\sim$1$\times$ to \textbf{2.2$\times$}; concurrency 16$\to$256 stays in the \textbf{1.3--1.8$\times$} range. Under variable prefix lengths ($\sigma$=0.0--0.6) PEEK's hit rate holds at 83--85\% while baselines degrade.

\paragraph{Takeaway.} Across both single- and multi-GPU offline serving, queue-derived prefix structure is the dominant lever for cache efficiency; the same insight that drives PEEK-online in this paper carries over to the batch regime, where the per-cycle pending tree collapses to a one-shot batch trie but the underlying co-design of prefix-grouped admission and queue-aware eviction is unchanged.

%% file: references.bib
@inproceedings{kwon2023vllm,
  title={Efficient Memory Management for Large Language Model Serving with {PagedAttention}},
  author={Kwon, Woosuk and Li, Zhuohan and Zhuang, Siyuan and Sheng, Ying and Zheng, Lianmin and Yu, Cody Hao and Gonzalez, Joseph and Zhang, Hao and Stoica, Ion},
  booktitle={Proceedings of the 29th Symposium on Operating Systems Principles (SOSP)},
  year={2023},
  xdoi={10.1145/3600006.3613165}
}

@inproceedings{zheng2024sglang,
  title={{SGLang}: Efficient Execution of Structured Language Model Programs},
  author={Zheng, Lianmin and Yin, Liangsheng and Xie, Zhiqiang and Sun, Chuyue and Huang, Jeff and Yu, Cody H. and Cao, Shiyi and Kozyrakis, Christos and Stoica, Ion and Gonzalez, Joseph E. and Barrett, Clark and Sheng, Ying},
  booktitle={Advances in Neural Information Processing Systems (NeurIPS)},
  year={2024},
  xurl={https://proceedings.neurips.cc/paper_files/paper/2024/hash/724be4472168f31ba1c9ac630f15dec8-Abstract-Conference.html}
}

@inproceedings{liu2024cachegen,
  title={{CacheGen}: {KV} Cache Compression and Streaming for Fast Large Language Model Serving},
  author={Liu, Yuhan and Li, Hanchen and Cheng, Yihua and Ray, Siddhant and Huang, Yuyang and Zhang, Qizheng and Du, Kuntai and Yao, Jiayi and Lu, Shan and Ananthanarayanan, Ganesh and Maire, Michael and Hoffmann, Henry and Holtzman, Ari and Jiang, Junchen},
  booktitle={Proceedings of the ACM SIGCOMM 2024 Conference},
  year={2024},
  xdoi={10.1145/3651890.3672274}
}

@inproceedings{gim2024prompt,
  title={Prompt Cache: Modular Attention Reuse for Low-Latency Inference},
  author={Gim, In and Chen, Guojun and Lee, Seung-seob and Sarda, Nikhil and Khandelwal, Anurag and Zhong, Lin},
  booktitle={Proceedings of Machine Learning and Systems (MLSys)},
  year={2024},
  xurl={https://proceedings.mlsys.org/paper_files/paper/2024/hash/a66caa1703fe34705a4368c3014c1966-Abstract-Conference.html}
}

@inproceedings{zhang2024h2o,
  title={{H2O}: Heavy-Hitter Oracle for Efficient Generative Inference of Large Language Models},
  author={Zhang, Zhenyu and Sheng, Ying and Zhou, Tianyi and Chen, Tianlong and Zheng, Lianmin and Cai, Ruisi and Song, Zhao and Tian, Yuandong and R{\'e}, Christopher and Barrett, Clark and Wang, Zhangyang and Chen, Beidi},
  booktitle={Advances in Neural Information Processing Systems (NeurIPS)},
  year={2023},
  xurl={https://proceedings.neurips.cc/paper_files/paper/2023/hash/6ceefa7b15572587b78ecfcebb2827f8-Abstract-Conference.html}
}

@article{jin2024ragcache,
  title={{RAGCache}: Efficient Knowledge Caching for Retrieval-Augmented Generation},
  author={Jin, Chao and Zhang, Zili and Jiang, Xuanlin and Liu, Fangyue and Liu, Xin and Liu, Xuanzhe and Jin, Xin},
  journal={ACM Transactions on Computer Systems (TOCS)},
  volume={44},
  number={1},
  pages={2:1--2:27},
  year={2025},
  xdoi={10.1145/3768628},
  venuecheck={https://arxiv.org/abs/2404.12457}
}

@inproceedings{ye2024chunkattention,
  title={{ChunkAttention}: Efficient Self-Attention with Prefix-Aware {KV} Cache and Two-Phase Partition},
  author={Ye, Lu and Tao, Ze and Huang, Yong and Li, Yang},
  booktitle={Proceedings of the 62nd Annual Meeting of the Association for Computational Linguistics (Volume 1: Long Papers)},
  year={2024},
  xdoi={10.18653/v1/2024.acl-long.623}
}

@inproceedings{liu2024scissorhands,
  title={Scissorhands: Exploiting the Persistence of Importance Hypothesis for {LLM} {KV} Cache Compression at Test Time},
  author={Liu, Zichang and Desai, Aditya and Liao, Fangshuo and Wang, Weitao and Xie, Victor and Xu, Zhaozhuo and Kyrillidis, Anastasios and Shrivastava, Anshumali},
  booktitle={Advances in Neural Information Processing Systems (NeurIPS)},
  year={2023},
  xurl={https://proceedings.neurips.cc/paper_files/paper/2023/hash/a452a7c6c463e4ae8fbdc614c6e983e6-Abstract-Conference.html}
}

@inproceedings{zhong2024distserve,
  title={{DistServe}: Disaggregating Prefill and Decoding for Goodput-optimized Large Language Model Serving},
  author={Zhong, Yinmin and Liu, Shengyu and Chen, Junda and Hu, Jianbo and Zhu, Yibo and Liu, Xuanzhe and Jin, Xin and Zhang, Hao},
  booktitle={18th USENIX Symposium on Operating Systems Design and Implementation (OSDI 24)},
  year={2024},
  xurl={https://www.usenix.org/conference/osdi24/presentation/zhong-yinmin}
}

@inproceedings{patel2024splitwise,
  title={{Splitwise}: Efficient Generative {LLM} Inference Using Phase Splitting},
  author={Patel, Pratyush and Choukse, Esha and Zhang, Chaojie and Shah, Aashaka and Goiri, {\'I}{\~n}igo and Maleki, Saeed and Bianchini, Ricardo},
  booktitle={2024 ACM/IEEE 51st Annual International Symposium on Computer Architecture (ISCA)},
  year={2024},
  xdoi={10.1109/ISCA59077.2024.00019}
}

@inproceedings{sun2024llumnix,
  title={{Llumnix}: Dynamic Scheduling for Large Language Model Serving},
  author={Sun, Biao and Huang, Ziming and Zhao, Hanyu and Xiao, Wencong and Zhang, Xinyi and Li, Yong and Lin, Wei},
  booktitle={18th USENIX Symposium on Operating Systems Design and Implementation (OSDI 24)},
  year={2024},
  xurl={https://www.usenix.org/conference/osdi24/presentation/sun-biao}
}

@inproceedings{yao2024cacheblend,
  title={{CacheBlend}: Fast Large Language Model Serving for {RAG} with Cached Knowledge Fusion},
  author={Yao, Jiayi and Li, Hanchen and Liu, Yuhan and Ray, Siddhant and Cheng, Yihua and Zhang, Qizheng and Du, Kuntai and Lu, Shan and Jiang, Junchen},
  booktitle={Proceedings of the Twentieth European Conference on Computer Systems (EuroSys)},
  year={2025},
  xdoi={10.1145/3689031.3696098}
}

@inproceedings{juravsky2024hydragen,
  title={{Hydragen}: High-Throughput {LLM} Inference with Shared Prefixes},
  author={Juravsky, Jordan and Brown, Bradley and Ehrlich, Ryan and Fu, Daniel Y. and R{\'e}, Christopher and Mirhoseini, Azalia},
  booktitle={ICML 2024 Workshop on Efficient Systems for Foundation Models II (ES-FoMo-II)},
  year={2024},
  xurl={https://openreview.net/forum?id=Ye3ia34Fpp}
}

@inproceedings{li2024snapkv,
  title={{SnapKV}: {LLM} Knows What You are Looking for Before Generation},
  author={Li, Yuhong and Huang, Yingbing and Yang, Bowen and Venkitesh, Bharat and Locatelli, Acyr and Ye, Hanchen and Cai, Tianle and Lewis, Patrick and Chen, Deming},
  booktitle={Advances in Neural Information Processing Systems (NeurIPS)},
  year={2024},
  xurl={https://proceedings.neurips.cc/paper_files/paper/2024/hash/28ab418242603e0f7323e54185d19bde-Abstract-Conference.html}
}

@inproceedings{yuan2024kivi,
  title={{KIVI}: A Tuning-Free Asymmetric 2bit Quantization for {KV} Cache},
  author={Liu, Zirui and Yuan, Jiayi and Jin, Hongye and Zhong, Shaochen and Xu, Zhaozhuo and Braverman, Vladimir and Chen, Beidi and Hu, Xia},
  booktitle={Proceedings of the 41st International Conference on Machine Learning (ICML)},
  year={2024},
  xurl={https://proceedings.mlr.press/v235/liu24bz.html}
}

@inproceedings{lee2024infinigen,
  title={{InfiniGen}: Efficient Generative Inference of Large Language Models with Dynamic {KV} Cache Management},
  author={Lee, Wonbeom and Lee, Jungi and Seo, Junghwan and Sim, Jaewoong},
  booktitle={18th USENIX Symposium on Operating Systems Design and Implementation (OSDI 24)},
  year={2024},
  xurl={https://www.usenix.org/conference/osdi24/presentation/lee}
}

@inproceedings{qin2024mooncake,
  title={{Mooncake}: Trading More Storage for Less Computation---A {KVCache}-centric Architecture for Serving {LLM} Chatbot},
  author={Qin, Ruoyu and Li, Zheming and He, Weiran and Cui, Jialei and Ren, Feng and Zhang, Mingxing and Wu, Yongwei and Zheng, Weimin and Xu, Xinran},
  booktitle={23rd USENIX Conference on File and Storage Technologies (FAST 25)},
  year={2025},
  xurl={https://www.usenix.org/conference/fast25/presentation/qin}
}

@inproceedings{gao2024cachedattention,
  title={Cost-Efficient Large Language Model Serving for Multi-turn Conversations with {CachedAttention}},
  author={Gao, Bin and He, Zhuomin and Sharma, Puru and Kang, Qingxuan and Jevdjic, Djordje and Deng, Junbo and Yang, Xingkun and Yu, Zhou and Zuo, Pengfei},
  booktitle={2024 USENIX Annual Technical Conference (USENIX ATC 24)},
  year={2024},
  xurl={https://www.usenix.org/conference/atc24/presentation/gao-bin-cost}
}

@inproceedings{srivatsa2024preble,
  title={{Preble}: Efficient Distributed Prompt Scheduling for {LLM} Serving},
  author={Srivatsa, Vikranth and He, Zijian and Abhyankar, Reyna and Li, Dongming and Zhang, Yiying},
  booktitle={Proceedings of the International Conference on Learning Representations (ICLR)},
  year={2025},
  xurl={https://openreview.net/forum?id=meKEKDhdnx}
}

@inproceedings{fu2024efficient,
  title={Efficient {LLM} Scheduling by Learning to Rank},
  author={Fu, Yichao and Zhu, Siqi and Su, Runlong and Qiao, Aurick and Stoica, Ion and Zhang, Hao},
  booktitle={Advances in Neural Information Processing Systems (NeurIPS)},
  year={2024},
  xurl={https://proceedings.neurips.cc/paper_files/paper/2024/hash/6c8985579293e0209bdaa4f21bb1d237-Abstract-Conference.html}
}

@article{cao2025dlpm,
  title={Locality-aware Fair Scheduling in {LLM} Serving},
  author={Cao, Shiyi and Wang, Yichuan and Mao, Ziming and Hsu, Pin-Lun and Yin, Liangsheng and Xia, Tian and Li, Dacheng and Liu, Shu and Zhang, Yineng and Zhou, Yang and Sheng, Ying and Gonzalez, Joseph and Stoica, Ion},
  journal={arXiv preprint arXiv:2501.14312},
  year={2025},
  xurl={https://arxiv.org/abs/2501.14312}
}

@inproceedings{dexter2025scheduling,
  title={{LLM} Query Scheduling with Prefix Reuse and Latency Constraints},
  author={Dexter, Gregory and Tang, Shao and Fatahi Baarzi, Ata and Song, Qingquan and Dharamsi, Tejas and Gupta, Aman},
  booktitle={Advances in Neural Information Processing Systems (NeurIPS)},
  year={2025},
  xurl={https://arxiv.org/abs/2502.04677},
  venuecheck={https://neurips.cc/virtual/2025/poster/118899}
}

@inproceedings{pan2025marconi,
  title={{Marconi}: Prefix Caching for the Era of Hybrid {LLMs}},
  author={Pan, Rui and Wang, Zhuang and Jia, Zhen and Karakus, Can and Zancato, Luca and Dao, Tri and Wang, Yida and Netravali, Ravi},
  booktitle={Proceedings of Machine Learning and Systems (MLSys)},
  year={2025},
  xurl={https://proceedings.mlsys.org/paper_files/paper/2025/hash/7c180af017258d239bac6248d1eb26ac-Abstract-Conference.html}
}

@article{li2025hotprefix,
  title={{HotPrefix}: Hotness-Aware {KV} Cache Scheduling for Efficient Prefix Sharing in {LLM} Inference Systems},
  author={Li, Yuhang and Gu, Rong and Huan, Chengying and Wang, Zhibin and Yao, Renjie and Tian, Chen and Chen, Guihai},
  journal={Proceedings of the ACM on Management of Data (PACMMOD)},
  volume={3},
  number={4},
  year={2025},
  xdoi={10.1145/3749168}
}

@inproceedings{lmcache2025,
  title={{LMCache}: An Efficient {KV} Cache Layer for Enterprise-Scale {LLM} Inference},
  author={Liu, Yuhan and Cheng, Yihua and Yao, Jiayi and An, Yuwei and Chen, Xiaokun and Feng, Shaoting and Huang, Yuyang and Shen, Samuel and Zhang, Rui and Du, Kuntai and Jiang, Junchen},
  booktitle={Proceedings of Machine Learning and Systems (MLSys)},
  year={2026},
  xurl={https://arxiv.org/abs/2510.09665}
}

@inproceedings{luo2025autellix,
  title={{Agentix}: An Efficient Serving Engine for {LLM} Agents as General Programs},
  author={Luo, Michael and Shi, Xiaoxiang and Cai, Colin and Zhang, Tianjun and Wong, Justin and Wang, Yichuan and Wang, Chi and Huang, Yanping and Chen, Zhifeng and Gonzalez, Joseph E. and Stoica, Ion},
  booktitle={23rd USENIX Symposium on Networked Systems Design and Implementation (NSDI 26)},
  year={2026},
  xurl={https://openreview.net/forum?id=LTofsJoTgf},
  venuecheck={https://arxiv.org/abs/2502.13965}
}

@inproceedings{hong2025sola,
  title={{SOLA}: Optimizing {SLO} Attainment for Large Language Model Serving with State-Aware Scheduling},
  author={Hong, Ke and Li, Xiuhong and Chen, Lufang and Mao, Qiuli and Dai, Guohao and Ning, Xuefei and Yan, Shengen and Liang, Yun and Wang, Yu},
  booktitle={Proceedings of Machine Learning and Systems (MLSys)},
  year={2025},
  xurl={https://proceedings.mlsys.org/paper_files/paper/2025/hash/bc82dbfbfa43232be85b8d9838f49c3e-Abstract-Conference.html}
}

@misc{llmd2025,
  title={{llm-d}: Kubernetes-native, High-Performance Distributed {LLM} Inference},
  author={{The llm-d Project}},
  year={2025},
  note={Open-source project website; the site does not list individual authors},
  xurl={https://llm-d.ai/}
}

@inproceedings{zhao2026blendserve,
  title={{BlendServe}: Optimizing Offline Inference with Resource-Aware Batching},
  author={Zhao, Yilong and Yang, Shuo and Zhu, Kan and Zheng, Lianmin and Kasikci, Baris and Qiao, Yifan and Zhou, Yang and Xing, Jiarong and Stoica, Ion},
  booktitle={Proceedings of the 31st ACM International Conference on Architectural Support for Programming Languages and Operating Systems (ASPLOS), Volume 2},
  year={2026},
  xdoi={10.1145/3779212.3790133}
}

@inproceedings{zheng2024batchllm,
  title={{BatchLLM}: Optimizing Large Batched {LLM} Inference with Global Prefix Sharing and Throughput-oriented Token Batching},
  author={Zheng, Zhen and Ji, Xin and Fang, Taosong and Zhou, Fanghao and Liu, Chuanjie and Peng, Gang},
  booktitle={Proceedings of Machine Learning and Systems (MLSys)},
  year={2026},
  xurl={https://openreview.net/forum?id=IuVHde07l6}
}

@inproceedings{pang2025kvflow,
  title={{KVFlow}: Efficient Prefix Caching for Accelerating {LLM}-Based Multi-Agent Workflows},
  author={Pan, Zaifeng and Patel, Ajjkumar Dahyalal and Hu, Zhengding and Shen, Yipeng and Guan, Yue and Li, Wan-Lu and Qin, Lianhui and Wang, Yida and Ding, Yufei},
  booktitle={Advances in Neural Information Processing Systems (NeurIPS)},
  year={2025},
  xurl={https://proceedings.neurips.cc/paper_files/paper/2025/hash/b7971d31a7d5eb0f1eed2f8f6f368195-Abstract-Conference.html}
}
